\documentclass[structabstract,longauth]{aa}  
\usepackage{graphicx}
\usepackage{rotating}
\usepackage{txfonts}
\usepackage{color}
\usepackage{natbib}
\usepackage{enumerate}

\begin{document}

 \title{The Gaia-ESO Survey: Target selection of open cluster stars\thanks{Based on data obtained with the European Southern Observatory telescopes under program 188.B-3002 (The $Gaia$-ESO Public Spectroscopic Survey).}}

\titlerunning{GES target selection in open clusters}
\authorrunning{A. Bragaglia et al.}

\author{
 A. Bragaglia\inst{1}, 
 E.~J. Alfaro\inst{2}, 
 E. Flaccomio\inst{3},  
 R. Blomme\inst{4}, 
 P. Donati\inst{1}, 
 M. Costado\inst{5},
 F. Damiani\inst{3},
 E. Franciosini\inst{6},
 L. Prisinzano\inst{3}, 
 S. Randich\inst{6},
 E.D. Friel\inst{7}, 
 D. Hatztidimitriou\inst{8},
 A. Vallenari\inst{9},
 A. Spagna\inst{10}, 
 L. Balaguer-Nunez\inst{11}, 
 R. Bonito\inst{3},  
 T. Cantat Gaudin\inst{11}, 
 L. Casamiquela\inst{12},
 R.~D. Jeffries\inst{13}, 
 C. Jordi\inst{14}, 
 L. Magrini\inst{6}, 
 J.~E. Drew\inst{15}, 
 R.~J. Jackson\inst{13},
 U. Abbas\inst{9},
 M. Caramazza\inst{16},  
 C. Hayes\inst{17}, 
 F.~M. Jim\'enez-Esteban\inst{18}, 
 P. Re Fiorentin\inst{9},
 N. Wright\inst{13},
 T. Bensby\inst{19},
 M. Bergemann\inst{20},
 G. Gilmore\inst{21}, 
 A. Gonneau\inst{21}, 
 U. Heiter\inst{22}, 
 A. Hourihane \inst{21},
 E. Pancino\inst{6,23},
 G. Sacco\inst{6},
 R. Smiljanic\inst{24},
 S. Zaggia\inst{9}
}

\institute{
INAF-Osservatorio di Astrofisica e Scienza dello Spazio di Bologna, via P. Gobetti 93/3, 40129, Bologna, Italy 
\and
Instituto de Astrof\'{i}sica de Andaluc\'{i}a-CSIC, Apdo. 3004, 18080, Granada, Spain
\and
INAF--Osservatorio Astronomico di Palermo, Piazza del Parlamento 1, 90134, Palermo, Italy
\and
Royal Observatory of Belgium, Ringlaan 3, 1180, Brussels, Belgium
\and
University of C\'adiz, C\'adiz, Spain
\and
INAF--Osservatorio Astrofisico di Arcetri, Largo E. Fermi, 5, 50125 Firenze, Italy
\and
Department of Astronomy, Indiana University, Bloomington, USA
\and
Section of Astrophysics, Astronomy and Mechanics, Department of Physics, National and Kapodistrian University of Athens, 15784, Athens, Greece
\and
INAF - Osservatorio Astronomico di Padova, Vicolo Osservatorio 5, I-35122 Padova,  Italy
\and
INAF - Osservatorio Astrofisico di Torino, Via Osservatorio 20, I-10025 Torino,  Italy
\and
Institut de Ci\`encies del Cosmos, Universitat de Barcelona (IEEC-ICCUB), Mart\'i i Franqu\`es 1, E-08028 Barcelona, Spain
\and
Laboratoire d’Astrophysique de Bordeaux, Univ. Bordeaux, CNRS, B18N, allée Geoffroy Saint-Hilaire, 33615 Pessac, France
\and
Astrophysics Group, Research Institute for the Environment, Physical Sciences and Applied Mathematics, Keele University, Keele, Staffordshire ST5 5BG, United Kingdom
\and
Dept. F\'\i sica Qu\`antica i Astrof\'\i sica, Institut de Ci\`encies del Cosmos (ICCUB), Universitat de Barcelona (IEEC-UB), Mart\'{\i} Franqu\`es 1, E08028 Barcelona, Spain
\and
Centre for Astrophysics Research, STRI, University of Hertfordshire, College Lane Campus, Hatfield AL10 9AB, United Kingdom
\and
Institut für Astronomie und Astrophysik Tübingen (IAAT), Sand, 1 72076 Tübingen Germany
\and
Department of Astronomy
University of Washington 
Box 351580, Seattle, WA 98195, USA.
\and
Departamento de Astrof\'{\i}sica,
Centro de Astrobiolog\'{\i}a (CSIC-INTA),  ESAC Campus, Camino Bajo del Castillo s/n, E-28692, Villanueva de la Ca\~nada, Madrid, Spain
\and
Lund Observatory, Department of Astronomy and Theoretical Physics, Box 43, SE-221 00 Lund, Sweden
\and
Max-Planck Institut f\"{u}r Astronomie, K\"{o}nigstuhl 17, 69117 Heidelberg, Germany
\and
Institute of Astronomy, University of Cambridge, Madingley Road, Cambridge CB3 0HA, United Kingdom
\and
Observational Astrophysics, Division of Astronomy and Space Physics, Department of Physics and Astronomy, Uppsala University, Box 516, SE-751 20 Uppsala, Sweden
\and
Space Science Data Center - Agenzia Spaziale Italiana, via del Politecnico, s.n.c., I-00133, Roma, Italy
\and
Nicolaus Copernicus Astronomical Center, Polish Academy of Sciences, ul. Bartycka 18, 00-716, Warsaw, Poland
}

\date{Received ; accepted }

  \abstract
{The $Gaia$-ESO Survey (GES) is a public, high-resolution spectroscopic survey conducted with FLAMES@VLT
from December 2011 to January 2018. GES targeted in particular a large sample of open clusters (OCs)  of all ages. }
{The different kinds of clusters and stars targeted in them are useful to
reach the main science goals,
which are the study of the OC structure and dynamics, the use of OCs to constrain and improve stellar evolution models,  and the definition
of  Galactic disc properties (e.g.  metallicity distribution). } 
{GES is organised in 19  working groups (WGs). We describe here the work of three of them, WG4  in charge of the selection of the targets
within each cluster), WG1 responsible for defining the most probable candidate
members, and WG6 in charge of the preparation of the  observations. As GES has been conducted before the second $Gaia$ data release, we could not make use of the $Gaia$ astrometry to define cluster member candidates. We made
use of public and private  photometry to select the stars to be observed with
FLAMES. Candidate target selection was
based on ground-based proper motions, radial velocities, and X-ray properties when appropriate, and it was mostly used to define the position of the clusters' evolutionary
sequences in the colour-magnitude diagrams. Targets for GIRAFFE were then selected  near the sequences in an
unbiased way. We used available  information on membership only for
the few stars to be observed with UVES.}
{We collected spectra for 62 confirmed clusters (and a few more were taken from the ESO archive). Among
them  are very young clusters, where the main targets are pre-main sequence
stars, clusters with very hot and massive stars currently on the main sequence,
intermediate-age and old clusters where evolved stars are the main
targets.  The selection of targets was as inclusive and unbiased as possible and we  observed a representative fraction of all possible targets, thus collecting 
the largest, most accurate, and most homogeneous spectroscopic data set on  ever achieved.} 
{}
\maketitle
%

\section{Introduction} \label{intro}
The $Gaia$-ESO Survey (GES, \citealt{gilmore12}, \citealt{r&g13}, and Gilmore et al. 2021, in prep., Randich et al. 2021, in prep.) is a large, public
spectroscopic survey    using the Fibre Large Array Multi Element Spectrograph (FLAMES)\footnote{FLAMES feeds two spectrographs, the high-resolution UVES and the low and intermediate-resolution GIRAFFE.}  instrument \citep{flames} on the European Southern Observatory (ESO) Very Large
Telescope (VLT-UT2) to obtain intermediate and high-resolution  spectroscopy of $\sim$10$^5$  stars in our Galaxy.  The observations
were conducted in the December 2011-January 2018 period, employing 340 nights. 
The goal of the $Gaia$-ESO Survey is to quantify the kinematical
and chemical abundance distributions  of the different components of the Milky Way, including the
bulge, thin and thick discs, halo, and a large sample of open clusters (OCs) that sample cluster age, mass, and distance well. We deal with open clusters in the present paper. Coming before the $Gaia$ mission results, the stars observed by $Gaia$-ESO were selected making use of several photometric sources, such as  the VISTA
Hemisphere Survey  \citep[VHS,][]{vhs},  the Skymapper project
\citep{skymapper},  and a  variety of photometric data for OCs (see below). 

As indicated by its name,  $Gaia$-ESO  intends to complement the
data from the $Gaia$ satellite\footnote{https://www.cosmos.esa.int/web/gaia/} (e.g. \citealt{mignard}, \citealt{prusti}), which was 
launched on December 2013. $Gaia$ is providing photometry, parallaxes, and proper motions of
exquisite quality for more than 1.5 billion objects, that is about 1\% of the  Galactic stellar
population. The first $Gaia$ data release (GDR1) happened on September 14, 2016 and contained information on the first 14 months of operation \citep[e.g.][]{GDR1a}. In GDR1 only positions and $G$ band photometry was released for about 1 billion sources. In addition, GDR1 provided parallaxes and mean proper motions for about 2 million bright stars in common with the Hipparcos and Tycho-2 catalogues - a realisation of the Tycho-$Gaia$ Astrometric Solution (TGAS). Although still limited,
GDR1 was used widely by the astronomic community and showed encouraging possibilities for open cluster studies, see for instance \citet{GDR1b} on TGAS astrometry of nearby clusters and a $Gaia$-ESO paper \citep{randichtgas} combining TGAS and $Gaia$-ESO data to improve the derivation of ages and comparison of stellar evolutionary models. 

$Gaia$ DR2 \citep{GDR2a}, published on April 25, 2018, contained positions, parallaxes, and proper motions for about 1.3 billion sources, together with photometry
    in $G$ (330-1050 nm), $G_{BP}$ (330-680 nm), and $G_{RP}$ (630-1050 nm) bands and radial velocities (RVs) for about 7 million sources \citep{GDR2c}. This catalogue brought about a revolution in Galactic studies \citep[see][to cite only one paper on Hertsprung Russell diagrams]{GDR2b} and made possible a more detailed analysis of the general  OC population and of membership in individual clusters, including new clusters discovered and candidate clusters removed, see for instance \citet{cantat18}, \citet{lp19}, \citet{sim19}, and \citet{castro20}.  

The third data release of $Gaia$ has been divided in two parts, and on December 3, 2020, EDR3 (i.e. early data release 3), published updated and more precise positions, parallaxes, proper motions, and photometry \citep{eDR3}, while the complete DR3, expected in the first half of 2022, will also comprise BP and RP spectra, classification and astrophysical parameters, etc (see https://www.cosmos.esa.int/web/gaia/release for details). $Gaia$ EDR3 has already been used to derive properties of stellar clusters \citep[e.g.][to cite only a case involving other space data, from the UVIT satellite]{jadhav21} and we expect a flourishing of studies larger than for DR2, especially once the full release will be available.

The $Gaia$ mission is also collecting spectroscopic observations with the RVS (Radial Velocity Spectrometer) at resolution $R\sim11500$ in a region near the Calcium {\sc ii} triplet (845–872 nm) to a
limiting magnitude of $G_{RVS}=16$. The final $Gaia$
DR is expected to provide RVs for about 150 million stars, with a precision strongly dependent on spectral type and magnitude. Only for the brighter stars will atmospheric parameters and element abundances be derived (for about five million stars brighter than 12, and two million stars brighter than 11, respectively).
$Gaia$ DR2 contained RVs for about 5 million FGK type stars down to $G_{RVS}=12$. For instance, \citet{soubiran18} used them to study the Galactic OC population; they were able to recover information on about 8000 stars in about 860 clusters, however, only 50\% of them had RV for at least three stars and 35\% had only one candidate member.
$Gaia$ DR3 will increase the sample to a few tens of millions stars.

While very useful, this means that there is a strong need for ground-based spectroscopic observations reaching fainter limits, at higher spectral resolution, and which can provide precise RVs and abundances.
The $Gaia$-ESO Survey will supplement the $Gaia$ RVS data for a significant subset of $Gaia$ targets, so that $Gaia$-precision
astrometry can be coupled with $Gaia$-ESO-precision RVs and chemical abundances. 

A large fraction of the $Gaia$-ESO programme is dedicated to the study of a 
large sample of OCs.    The  top-level scientific goals of the cluster
component of $Gaia$-ESO are described in Randich et al. (2021) and what follows is a brief summary.

Firstly, we propose to understand how clusters form, evolve, and  eventually dissolve and disperse, through the investigation of internal cluster kinematics and dynamics. In fact,  clusters may contribute most
of the stars of the Milky Way field and are valuable tools for the study of the formation and evolution  of the Galactic disc.

Secondly, we intend to pursue the calibration of the complex
physics involved in stellar evolution, using clusters as templates at
different age, mass, and chemical composition. In fact, to a first approximation, OCs are  observational isochrones.

Thirdly, we aim at obtaining the detailed study of the properties  and evolution of the  Milky Way thin disc. This is achieved through the study of the distribution of chemical abundances and of their evolution with time. 

The FLAMES spectra allow us to determine RVs for all observed
stars; this  permits  the identification of true cluster members in all evolutionary
phases, from the pre-main sequence (PMS) to the evolved giants, to be used as
observational templates for stellar  evolution theory.  In nearby clusters
(within about 1.5 kpc, \citealt{jackson15}) the precision reached in RV (down to 0.25 km~s$^{-1}$) is  sufficient to resolve the internal velocity dispersion and give a measure of the internal
kinematics \citep[see e.g.][]{jeffriesGES}.  This is especially important when coupled with the precise
positions, distance, and proper  motions from the $Gaia$ satellite \citep[see e.g.][the  first paper to combine $Gaia$-ESO RVs and $Gaia$ PMs for a large-scale kinematic study of a young cluster]{wright19}. The $Gaia$-ESO
spectra also provide metallicity and detailed chemical abundances
for OCs that sample cluster age, mass, position, and distance well. 
Those data are fundamental for the study of the metallicity distribution in the disc and its evolution with time, thus providing key input to the chemical evolution models of the Milky Way
disc.

A detailed description of the survey, of its goals and methods can be found
in the papers by Gilmore et al. and Randich et al.  (2021, both in preparation). We recall here that the $Gaia$-ESO observations are obtained with both the
GIRAFFE spectrograph (about 130 fibres, with resolution $R\sim15000-25000$,
depending on the setup used, with a wavelength coverage of a few tens of nm),
and the UVES spectrograph (6 or 8 fibres, depending on the setup, with $R\simeq47000$,
and covering about 200 nm).  Table~\ref{tab1} gives a summary of the gratings
used, their characteristics,  the kind of clusters (and stars within clusters)
that are observed with them, and the most important lines and elements
visible with each setup. Some OCs, used as calibrators, were also observed with the same setups of the field stars (HR10, HR21) of the Galactic survey; for them, the selection of stars followed a different method and details can be found in \citet{pancino17}.

We focus here  on the target selection process for the open clusters and
on the observation's preparation. In particular, Secs~2 and 3 briefly describe the selection of clusters and the kind of stars targeted in each cluster, respectively. Section~4 deals with the creation of the catalogues of stars to observe.  Section~5  presents the work-flow of the process and the actual preparation of observations. A discussion
on the fraction of actual members observed, based also on posterior $Gaia$ data, is done in Sec.~6.
A short summary is given in Sec.~7.

\begin{table*}
\centering
\setlength{\tabcolsep}{1.1mm}
\caption{Setups used for GIRAFFE and UVES observation of open clusters.}
\begin{tabular}{lccccll}
\hline
Setup & $\lambda\lambda$ & $R1$ & $R2$&No    & Stars and clusters on which &Prominent lines and elements\\
      & (nm)             &     &    &fibres   & the setup is mainly used                        &\\
\hline
\multicolumn{6}{c}{UVES}\\
520   &	414-621 & 47000 & & 6 & early type stars & H$\gamma,\beta$; and see below the case of blue setups \\
580   & 476-684 & 47000 & & 8 &late type stars & H$\alpha$; Fe {\sc i}, {\sc ii}; Fe-peak; $\alpha$-elements; Na; [O {\sc i}]; Al; n-capture; Li\\
\multicolumn{6}{c}{GIRAFFE}\\
HR03   & 403.3-420.1 &24800 &31400 &130 & early-type st., massive young cl.  
 & H$\delta$; He {\sc i}; Si {\sc ii}; {\sc iv}; O {\sc ii}; [Si {\sc ii}]\\  
HR04   & 418.8-439.2 &    & 24000 &130 & early-type st., massive young cl.  
 & H$\gamma$; He {\sc i}; He {\sc ii}; Si {\sc iv}; N {\sc ii}\\     
HR05a  & 434.0-458.7 &18470 &20250 &130 & early-type st., massive young cl.  
 & He {\sc i}, {\sc ii}, Si {\sc iii}; Mg {\sc ii}; N {\sc ii}; {\sc iii}; O {\sc ii}  \\   
HR06   & 453.8-475.9 &20350 &24300 &130 & early-type st., massive young cl.   
 & He {\sc i}, {\sc ii}; Si {\sc iv}; C {\sc iii}; N {\sc ii} and N {\sc iii}; O {\sc ii} \\   
HR09b  & 514.3-535.6 &25900 &31750 &130 & early-type st.   & Mg b; Fe {\sc i}, {\sc ii}; Ti {\sc ii}; Cr {\sc i}, {\sc ii} S {\sc ii}; Mn {\sc ii} \\   
HR14a & 630.8-670.1 &17740 &18000 &130 & early-type st., massive young cl. 
 & H$\alpha$; He {\sc i}, {\sc ii}; Si {\sc ii}; C {\sc ii}; Ti {\sc i}; Ba {\sc ii}; TiO\\   
HR15n & 647.0-679.0 &17000 &19200 &130 & late-type st., all cl.   & H$\alpha$; He {\sc i},{\sc ii}, Li; Fe {\sc i}; Ca; Si; Mg; Ti {\sc i}; Ba {\sc ii}; [S {\sc ii}]; [N {\sc ii}]; TiO; CaH\\   
\hline
\end{tabular}
\begin{list}{}{}
\item[-] $R1$, $R2$ are the resolution before and after the GIRAFFE upgrade in February 2015, respectively (no change for UVES). The HR04 setup was used only after the upgrade.
\item[-] The number of allocated stars are less than the fibres, since some ($\ge1$ for UVES, $\gtrsim$15 for GIRAFFE) are dedicated to sky positions. 
\item[-] We define here stars of O, B, and A spectral type as `early' and stars
of F, G, K, and M spectral type as `late'. We note that some A-type stars have been observed  using UVES 580.
\item[-] For a few clusters used as cross-survey calibrators, also HR10 (533.9-561.9 nm) and HR21 (848.4-900.1 nm) exposures of a fraction of the targets are acquired.
\end{list}
\label{tab1}
\end{table*}

\section{The selection of clusters} \label{sec2}

To reach the top-level science  goals mentioned above, the $Gaia$-ESO Survey
targets a very  large sample of clusters, covering the whole
age-metallicity-mass-Galactic location-density  parameter space. Within each
cluster, we observe a large and unbiased sample of  stars using the
GIRAFFE fibres and a smaller, biased sample of the most likely cluster members 
using the UVES fibres. While RV, atmospheric parameters, and
metallicity are obtained for the entire sample of  clusters and targets within them, there
are necessarily also some differences in the way categories of clusters  are
dealt with, due to the large variety of properties among  OCs. These
differences are reflected in the selection of targets and the choice of the
gratings. However,  this does not imply divergence of goals; on the contrary, we
build on this variety to obtain a comprehensive picture of the open clusters'
family and of their importance for understanding the Milky Way.

The full description of the cluster selection  will appear in Randich et al. (2021).  
We  recall here only a few concepts, relevant for the target selection within
each cluster.  We can divide the $Gaia$-ESO clusters into two main classes and two
sub-classes each.

Firstly, we have young clusters (age $\leq$100~Myr). They may be without or with massive stars
(mass $\ge 8$ M$_\odot$).

Secondly, we have intermediate-age and old clusters (age $>100$ Myr, up to several Gyr\footnote{About 10\% of all OCs observed have age larger than 4 Gyr; while this could be defined intermediate-age, these clusters are usually referred to as old in OC literature.}). They may be
without or with a red clump (RC).

The two groups are about one third and two thirds of all observed clusters (see Table~\ref{t:app}). This rather artificial division is however useful to characterise the kinds of stars that dominate in the observed clusters and what information can be best extracted from each category to fulfil the survey main goals.
 
\paragraph{{\em 1a)} Young open clusters, with no or few massive stars.} This group, dominated by a late-type population with at most a few early-type stars,
includes young clusters and associations just out of the embedded phase and up to an
age of 100 Myr. We focus on clusters in the solar vicinity, up to about 1.5 kpc,
to enhance the $Gaia$ connection and to be able to reach low-mass cluster members
(see below).  These clusters are crucial systems to understand the ongoing  star
formation (SF) processes  and the recent SF history of the Galaxy. These processes involve
stellar and dynamical evolution, chemical enrichment mechanisms,  and connection
with the surrounding environment.  Data from the $Gaia$ ESO survey have brought about significant progress in
these studies, especially for the low-mass populations of these systems, thanks to different
spectroscopic diagnostics. Radial velocities, presence of at least one between strong lithium absorption (indication of
youth) and H$\alpha$ in emission (indication of  activity or accretion
processes), and gravity-sensitive spectral lines are all effective diagnostic of cluster membership. The main effort, in young clusters and associations, is to obtain a complete and unbiased sample of cluster stars, for which precise RVs -and possibly abundances- are measured; our strategy is then driven by the GIRAFFE observations.  By coupling proper motion data from $Gaia$ and $Gaia$-ESO accurate RVs, we are able to determine the cluster dynamics.

Comparison of clusters of different age, environment, structure, and morphology will allow us to understand  the dynamical evolution with time of such systems. The data can also help to confirm/refute the claims of triggered star formation in some star forming regions (SFR). Finally, the $Gaia$-ESO spectra, especially those obtained with UVES, provide chemical abundance diagnostics, contributing  important constraints on the metallicity distribution evolution.

\paragraph{{\em 1b)}   Massive-star young clusters.} They complete the cluster
parameter space covered by the  young clusters; as they are rare, they are necessarily observed also up to much
larger distances than clusters discussed in point 1a.  We concentrate on
studying the population of young, massive, and hot main sequence (MS) stars and their influence on
the clusters. The massive stars profoundly affect the evolution of the  cluster
as they shed large amounts of mass, momentum, and energy (ionising radiation), which  may lead to the
dispersal of the parental molecular cloud, and hence to the end  of the star
formation in that cluster \citep{lada}. Knowledge of the kinematics of the
massive stars and, when feasible, of the lower-mass members of the same cluster, is highly relevant to the cluster dynamics. They are also
interesting objects in themselves as they put important constraints on stellar
evolution models. 
  
In addition, massive stars are important in determining Galactic abundance  gradients
\citep{daflon}. Their high luminosity allows us to  cover larger distances in
the Galaxy. As they are very young, their  abundance values are much closer to
the present-day ones. From a comparison  between the Galactocentric abundance gradient of young
and older clusters one can derive the  time evolution of the gradient, leading to a
powerful constraint on models  of thin disc formation.

\paragraph{{\em 2)} Intermediate-age and old clusters, with or without red
clump.}   The RC is the locus of stars burning He in their core and
is visible in clusters older than about 300-400 Myr, in which a
(conspicuous) number of stars evolved from the MS are present.  The old OCs
are valuable tools to study the formation  and evolution of the Galactic
disc and  rare fossils of its past star formation history. The vast majority of 
stars born in OCs are indeed dispersed into the Galactic field in a relatively short time (e.g. \citealt{janes94}, \citealt{gieles06}), 
and thus old survived clusters are unique relics of the composition of the interstellar
medium (ISM) at the epoch of their formation \citep[e.g.][]{friel95}.  The
study of a well defined sample of clusters in terms of age and Galactocentric 
distance allows us to understand the spatial distribution of elements in the Milky Way
disc and to investigate  its evolution with time. The main focus for these
open clusters is on the determination of precise chemical  abundances, so emphasis is on the
spectra of the relatively few stars observed with the UVES fibres.  The UVES
targets are chosen  preferentially from stars in the RC when it is present, to ensure the
best homogeneity among different clusters, since they are stars in a well constrained evolutionary phase and span a small range in atmospheric parameters. A secondary goal is to define cluster
membership using the RVs obtained with the more numerous GIRAFFE fibres. A clean
definition of the evolutionary sequences and especially of key features, such as 
the MS turn-off, the sub-giant branch (SGB), the red giant branch (RGB), and the
RC, is fundamental to derive age through fits to theoretical isochrones. In fact, 
$Gaia$ provides us with very precise parallaxes, so distance can be derived, but age can ultimately be obtained only through comparison with theoretical models.
Cluster distance is essential to define the radial distribution of metallicity
and chemical abundances in the Galaxy, while precise and homogeneous ages are
required to investigate the chemical evolution history.  Finally, whenever
sufficiently cool spectral types are present on the MS, then lithium abundances
can be measured also from the more numerous GIRAFFE spectra.

\smallskip
This division among clusters of different types is of course only a scheme, and for instance, internal dynamics can be
studied also for old and  nearby clusters. Important legacy items are the
definition of cluster members, the study of the initial mass function in nearby
clusters, of the lithium depletion, and of the chromospheric activity. Finally, $Gaia$-ESO obtains  spectra for
PMS, MS stars of all mass and spectral types, and evolved giants. 
This very large sample of stars at all evolutionary phases in clusters of
different ages and  chemical compositions will give important and stringent
constraints to stellar evolutionary models.

\begin{figure*}
\centering
\includegraphics[bb=1 190 600 530,clip,scale=0.85]{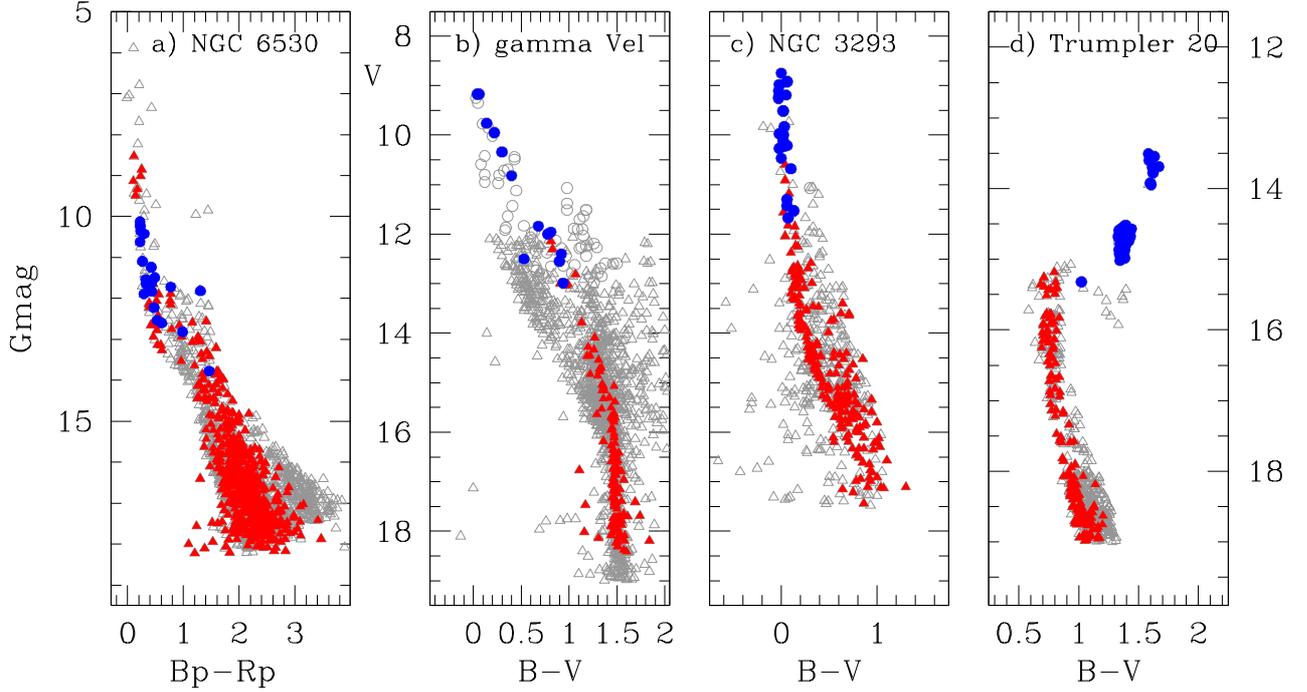}
\caption{Examples of targets in four clusters of different ages:
(a) NGC~6530; (b) $\gamma$~Vel (see text for the numerous field stars), (c) NGC~3293, and (d) Trumpler~20; their age is about 2 Myr, 20 Myr, 10 Myr, and 1.8 yr (see Table~\ref{t:app}). Open grey symbols indicate all stars
observed (triangles GIRAFFE, circles UVES); filled symbols indicate
candidate members (red triangles GIRAFFE, blue circles UVES) according to $Gaia$-ESO data. 
}
\label{fig4oc}
\end{figure*}

\section{Targets observed in the different kind of clusters} \label{whichtargets}

Before delving into a detailed description of target selection and observation preparation, we briefly recall the role of the WGs involved. All $Gaia$-ESO activities are organised in WGs and cluster stars observation preparation is done by three of them: WG1 (cluster membership analysis, led by E. Alfaro), WG4 (cluster target selection, led by A. Bragaglia), and WG6b (FPOSS/OB\footnote{FPOSS means Fibre Positioner Observation Support Software and it is the fibre configuration programme for the preparation of FLAMES observations; OB stands for observing block, that is the set of instructions for observation execution.} generation, led by E. Flaccomio); the WG2 activities (auxiliary data for cluster target selection) were merged with WG1 and WG4. For more details on the WG structure and operations, see Gilmore et al. (2021).

The science goals and immediate objectives, as described in Sects~\ref{intro}
and \ref{sec2},  drive the  star selection in each cluster. The target stars are
selected from the colour-magnitude diagrams (CMDs, see Sects~\ref{emilio} and
\ref{workflow}), taking into  account both the position of each star with
respect to the cluster evolutionary sequences and the distance from  the cluster
centre (the latter only if the cluster does not fill the entire FLAMES field of view - FoV). The actual description of the targets selection will be detailed in the next
section, we concentrate here on the kind of targets we are dealing with.

We remark that the (few) UVES targets are generally selected from the most
secure members, on the basis of the available auxiliary information such as previously published RVs, proper
motions, X-ray properties, lithium abundance, chromospheric activity,  depending
on the cluster type. For cases where no such auxiliary information was available, as for instance in 25~Ori, 
we were forced to observe less secure members. In the case of old clusters, the main targets for UVES are
RC stars, while RGB stars have second priority. In
intermediate-age clusters, MS stars may be targeted as well. For young clusters,
UVES targets are PMS  and MS stars.  In addition to the main targets, for
nearby clusters of all ages we try to observe (also) some MS stars of late spectral type, for completeness and
cross comparison with GIRAFFE. The faint magnitude limit for the UVES targets is $V\simeq16.5$, which is the limit to
obtain S/N$\sim$50 in six to seven hours\footnote{The exposure times for UVES and GIRAFFE
were decided in order to reach the
intended scientific goals in the allotted time, see Randich et al. (2021) for a full
justification.}. 

The selection of GIRAFFE targets is aimed at observing inclusive and unbiased
samples of cluster star candidates rather than only high probability members (see next section).
While we aim at a high degree of  statistical completeness, only a significant sub-sample of
candidate members is observed in very rich or  extended clusters, to avoid
excessive use of telescope time. Targets are PMS or MS stars (and  evolved
stars in old OCs, to ensure RV membership determination for stars in all evolutionary phases), with $V\le19$, to match the $Gaia$ mission good astrometric precision at the faint limit. For instance, in $Gaia$ DR2 
the median uncertainty in parallax (proper motions) is about 0.04 (0.05) mas~yr$^{-1}$ for $G < 14$ mag sources, 0.1 (0.2) mas~yr$^{-1}$ at $G = 17$ mag, and 0.7 (1.2) mas~yr$^{-1}$ at $G = 20$ mag \citep{GDR2d}. These values are already better in EDR3 and their precision will increase in further data releases based on longer time scales or the full mission duration.
In some clusters, the faint limit for GIRAFFE targets could be  extended to $V \simeq 19.5$ in order to utilise otherwise spare fibres.

According to the division of clusters' type given in Sect.~\ref{sec2}, the
following kinds of stars are observed:

\begin{itemize}
\item[{\em 1a)}] In young clusters and SFRs without  a
dominant population of early type stars, the targets for GIRAFFE are late-type
(F to M) stars in the magnitude range 12$\leq$V$\leq$19, in the PMS or MS
phase.  They are observed with the setup HR15n (containing H$\alpha$ and the Li
670.7~nm line, both important diagnostic lines). The UVES targets are chosen in the magnitude range  9$<$V$<$15. They  are observed with the
580nm setup if of late spectral type, and with the 520nm
setup if (a few) bright, early-type stars are targeted. 
If the information is available,  the UVES targets are preferentially selected to be slow rotators ($v\sin i <$ 15 km~s$^{-1}$) and not strong accretors ($dM/dt< $ 10$^{-10}$ M$_{\odot}$ yr$^{-1}$).

\item[{\em 1b)}] In young clusters dominated by massive early-type stars, the
targets observed with GIRAFFE are  B and A-type stars, down to $V\simeq18$. They
are observed with the blue setups HR03, HR04, HR05a, HR06, HR09b, and HR14a
(containing H$\alpha$). The UVES fibres are allocated to O-type stars in the
magnitude range $V=9-15$ and the 520nm setup is used.

\item[{\em 2a)}] In old and intermediate-age clusters with RC stars,  the
GIRAFFE targets are mainly stars from the MS turn-off (TO) down to V$=19$. They are
observed with HR09b if the spectral type is A to F and with HR15n for later
types. Giant stars may also be targeted, with HR15n. Stars observed with UVES (always with the 580nm setup) are preferentially
on the RC. We may  also observe RGB stars, as previously mentioned,  if  the RC is too faint because of distance or extinction, or if there are only a few RC stars.   In some
nearby clusters, FGK MS stars are also observed with UVES, to compare results with giants. For
the UVES targets in the old clusters the magnitude range is $V=9-16.5$; the
fainter limit for these clusters is determined by the requirement to study the
metallicity distribution in the disc, hence they are necessarily distributed 
also out to larger distances. 

\item[{\em 2b)}] In intermediate age open clusters without RC stars (i.e. with
age between 100 and 300-400 Myr), the  target selection is the same as for case {\em 1a)}. The choice of setups for GIRAFFE  targets is as for case {\em 2a)}.

\item[{\em 3)}] Whenever possible, a few stars are repeatedly observed in different configurations (e.g. with UVES and GIRAFFE, or in different pointings, or with two or more GIRAFFE setups) to perform sanity checks on the quality of the derived parameters.

\end{itemize} 

These different choices are illustrated in Fig.~\ref{fig4oc} for four clusters observed early during survey  operation, a) one very young cluster where we targeted both low-mass and early-type stars (NGC~6530, see \citealt{prisinzano19}), b) one young
cluster where we did not target early-type stars ($\gamma$~Vel, age 5-10 Myr, see \citealt{jeffries} and \citealt{jeffriesGES}, \citealt{spina} for
first results from the $Gaia$-ESO Survey, or 20 Myr, see Table~\ref{t:app}),
c) one with many massive, early-type stars (NGC~3293, age about 10 Myr, see
\citealt{baume}, \citealt{delgado}), and d) one old cluster (Trumpler~20, age about 1.5 Gyr, see for example \citealt{carraro} and for first $Gaia$-ESO results, \citealt{donatitr20}). 
In NGC~6530 we targeted almost 2000 stars, of which 55 with UVES 520nm or 580nm (for a total of 661 candidate members, see \citealt{prisinzano19}).
In $\gamma$~Vel we observed about 1240 stars with GIRAFFE (135 are candidate members)
and 80 stars with UVES (13 are candidate members); in NGC~3293 we observed
540 stars with GIRAFFE (210 members) and 26 with UVES (25 members); in Trumpler~20
we observed 525 stars with GIRAFFE (156 members) and 42 with UVES (41 members). The number of member stars is
based on RVs (and other indicators for $\gamma$~Vel) and is taken from the first analysis of the data; it may vary a little with a more in-deep procedure taking into account, for instance, also the presence of binaries and astrometric information (see Sec.~\ref{members}), but is accurate
enough to illustrate the situation.
The apparently low success rate in $\gamma$~Vel depends on the fact that we preferred to assign fibres to low priority targets rather than leave them unused; even so, the observations brought very interesting results, for instance demonstrating the existence of two sub-groups, $\gamma$~Vel A and B \citep{jeffriesGES}. In any case, the spectra of field stars may be useful as a legacy, especially now that their distance is known thanks to the $Gaia$ satellite results, see e.g. \citet{magrini21} and \citet{romano21}, where field stars observed in the same field as the targeted OC were used to study the lithium  distribution and evolution in giant stars.

\begin{figure*}
\begin{center}
\includegraphics[scale=0.8,bb=20 180 580 600, clip]{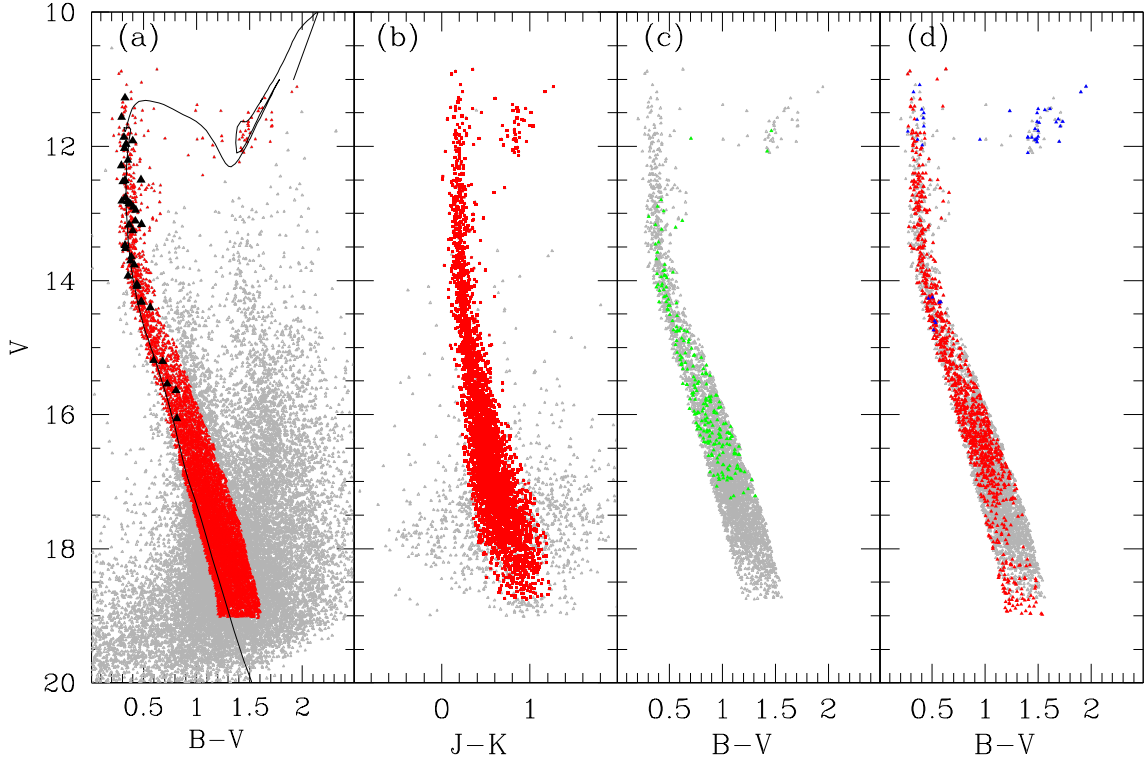}
\caption{Step by step procedure (from left to right) applied in the
selection of candidate members. (a) $V,B-V$ CMD for  NGC\,6705, showing the selection of candidate
members. In grey we show all stars with optical photometry, while  red filled triangles
represent the stars selected on the basis of $BV$ photometric information. The
line is the best-fit PARSEC \citep{parsec} isochrone \citep[see][]{cantatm11},
while the black symbols are selected members from kinematic analysis (see
section~\ref{emilio}). (b) The stars selected in the optical bands are shown in
grey, in the $V$ vs. $J-K$ CMD; here red points represent the stars selected
based on $JHK$ analysis. (c): Candidate targets in grey, with kinematic outliers
shown in green. (d) Final selection  (in grey) is shown together with the
stars actually observed (in red with GIRAFFE, in blue with UVES fibres).
}
\label{ngc6705_fig1}
\end{center}
\end{figure*}

\section{Tailoring a primary list of tentative star targets}\label{emilio}

As has already been said in the previous sections, the strategies followed for
selecting UVES and GIRAFFE targets are very different, since they answer two
different questions.  The (few) UVES spectra are used to derive the metallicity
and detailed abundances and efforts are made to assign the few fibres to the most
secure member stars. This means, depending on the cluster type, that we considered
auxiliary literature information: RV and proper motions (for the older
clusters) and additionally X-ray properties, lithium  abundance, H$\alpha$ emission, 
chromospheric activity, and rotation (for the younger clusters and the SFRs). The
choice of UVES targets is then strongly biased and will not be discussed anymore
in this section.

The (many) GIRAFFE spectra are used to study the general properties of the
clusters and the following description applies to the selection of GIRAFFE
targets. Since one of the fundamental aspects of the $Gaia$-ESO project is its legacy 
character, the selection of targets must produce a catalogue  representative of
a sample as much complete and unbiased  as possible, which is also potentially useful and easily accessible to the astronomical community interested in  similar
scientific objectives or which can be used beyond the original survey goals (this is of course valid also for the UVES spectra). While we aim at reaching a complete  coverage of the
clusters, practical considerations on the time required force us to  limit 
ourselves to a representative sample. 
There are three main constraints we considered for tailoring the primary list of target stars.

Firstly, we aim at homogeneity. The procedure has to be as homogeneous as
possible even if the  data are not and the clusters under study have very
different observational characteristics, such as  different magnitude ranges,
different degrees of field star contamination, different  photometric systems,
existence of RV data or good-quality proper motions, etc.

Secondly, we require simplicity. The tools used should be  easily
accessible for anyone, in such a  way that the verification and control of the
final product can be reproduced by anyone.

Thirdly, we need unbiased samples. Only those stars considered with
certainty as non-members  should be excluded from the final list (i.e. the
evident outliers from the stellar population  of the  cluster). Although this
criterion entails the possible inclusion of (a significant fraction of) field
stars, the  final list will be more in line with the general objectives of the
project and its character as a legacy  programme than if we performed a more
restrictive purge.  

Therefore, for GIRAFFE targets we selected candidate cluster members on the basis of photometry. Other astronomical data  were only used to define the spatial
extent and the evolutionary sequences for each cluster, along which we picked stars to be observed. Furthermore, proper motions were used, when possible,  to exclude a small fraction of potential targets visualised as kinematic outliers (see the description below).

We describe in this section the data sources, the tools, and the kinematical
selection of members  that helped the final selection. We present details for
one cluster as example. 
 
\subsection{Data sources}\label{datasources}

The target selection involves a laborious process of looking for potential sources of optical photometry (and ancillary information on membership). This means that only clusters where there are sufficient photometric data are included in our sample. 
 
Photometry and kinematic information on the cluster sample are the basic
ingredients for the target  selection. Unfortunately, no single, homogeneous, and
all-sky dataset covers all the  clusters of the $Gaia$-ESO programme,\footnote{We use the
Two Micron All Sky Survey \cite [2MASS,][]{2mass}, as described in the text, but
it does not reach faint enough and with the required  precision. Furthermore,  $JHK$ data alone are insufficient to select a complete sample of cluster 
members without including far too many non-members.} so we resorted to individual available photometric studies. 
However,  the European Galactic Plane Surveys (EGAPS)\footnote{http://www.ing.iac.es/Astronomy/development/iphas/ \\
http://www.vphasplus.org }  optical photometry covers part of the cluster sample:
IPHAS and UVEX (INT Photometric H$\alpha$ Survey, \citealt{iphas}, UV-Excess Survey, \citealt{groot}) in the north  and the public ESO survey VPHAS+ (VST Photometric H$\alpha$ Survey, \citealt{vphas}) in the south.  Collectively, these surveys use the Sloan {\sl ugri} broadband filters, and H$\alpha$
narrow band, achieving a 5$\sigma$ faint limit $>20$ in all bands.  The bright limit is typically 12-13th magnitude.  This suits them well to our selection needs.

Individual CCD optical photometric studies were used for some of the
clusters; in some cases, they were the only source of targets.  The public data
were retrieved from  the WEBDA\footnote{http://webda.physics.muni.cz/} database
and the VizieR catalogue access tool\footnote{https://vizier.cds.unistra.fr/viz-bin/VizieR}. In a few cases, specific photometric
studies have been performed for selected clusters  or archive wide-field images
have been analysed. Details on the photometric source(s) are presented in the various $Gaia$-ESO papers on OCs
(e.g. \citealt{donati3oc}, for
Berkeley~81 and S. Zaggia, private communication, for NGC~4815 and NGC~6705). 

In the near-infrared range, the 2MASS catalogue \citep{2mass}  is the main data
source. These data are used for membership selection and for establishing an
astrometric reference frame for fibre positioning. The 2MASS catalogue, given
its all-sky character and its well-proven accurate astrometry (better than 0.1
arcsecond for sky positions), is the basis of our
pre-$Gaia$ coordinate system.

Several ground-based, large-area proper motion (PM) catalogues have been the source of our 2D
kinematic information: the  UCAC4 \citep[Fourth US Naval Observatory CCD Astrograph
Catalogue,][]{zacharias}, SPM4 \citep[Southern Proper Motion Program
IV,][]{girard},   and PPMXL \citep[Positions and Proper
Motions-Extended,][]{roeser} catalogues.
When available, we used preferentially the UCAC4 all-sky catalogue or the SPM4 PM data.   Table~\ref{astrocat} gives a summary of the main catalogues and  other physical variables involved in the  targets' selection; for specific clusters, RV data and other  physical information were  taken from the ViZieR and WEBDA databases. 

\begin{table}{}
 \caption{Summary of main catalogues used in the target selection.}
  \begin{center}
\begin{tabular}{ c   c   c   c  }
\hline 
Optical & NIR   & PM  &  Others \\  
\hline 
EGAPS & 2MASS & UCAC4   & Li abundance \\
IPHAS/UVEX & & SPM4  & X-ray \\ 
VPHAS & & PPMXL &  RV \\
\hline
\end{tabular}
\end{center}
\label{astrocat}
\end{table}

Literature data concerning RVs for cluster stars are more sparse, non-homogeneous, and far
from complete for the cluster selection.  The WEBDA and the catalogues at
CDS are the best RV data providers.   Also in this case, some effort has been
devoted to obtaining new spectra to measure RVs for stars in $Gaia$-ESO clusters \citep[e.g.][]{hayes_friel}. 

Finally, other physical variables, such as metallicity, X-ray emission, Li
abundance, and any other information that might  help discriminate between
cluster members and field stars was used to help define the cluster photometric
sequences.  In particular, X-ray emission and lithium  abundance are useful for young
clusters (age less than about 100 Myr) and SFRs. In these cases
we identified candidate members based on their young age with respect to older,
disc stars. In fact, young stars show X-ray fluxes significantly larger than
those observed in older stars of the same spectral type
(e.g. \citealt{favata_micela}, \citealt{feigelson}). 
Lithium is easily destroyed at relatively low
temperatures, starting already in PMS. Thus, presence of high lithium abundance is a well known tracer of youth and indicates a high probability of belonging to the SFR under study (see e.g. the analysis in a few $Gaia$-ESO papers on young clusters: \citealt{jeffriesGES}, \citealt{rigliaco16}, \citealt{sacco17}, and \citealt{wright19}, all of which rely on the presence of lithium to determine youth for studies of these clusters).

\subsection{Tools and Procedure}\label{methodology}

The selection of the candidate cluster members was primarily based  upon their
position in the CMDs. This means we took into account the expected location of the different evolutionary
sequences in the various observational  diagrams. 

The first step is the definition of the cluster sequences and the spatial extent
of the cluster. For this we made use of both the theoretical isochrones that
best fit the  published physical  parameters of the cluster and the location of
those stars that have membership information. We used the PARSEC  isochrone models (PAdova and TRieste Stellar Evolution Code, \citealt{parsec}) for the MS and post-MS stars of the clusters in any age range and those calculated by \citet{siess} for the PMS stars. 

Then we selected the regions in the CMDs around the cluster main loci. The width of the CMD strip used for selection depends on several factors, such as variable reddening, binary sequence, and photometric errors, but always taking into account  that we need to be inclusive. The
subjective component of this selection is moderated by the use of multi band
photometry, with the possibility of analysing various CMDs to better define the sequences.

The next step was then assigning a new variable to each star, indicating whether it
is a candidate member or not, according to its position in the various CMDs,
independently in the optical and near-IR bands. The 2MASS photometry
does not always reach as deep as the optical one, so the latter is more important for
the fainter part of the sequences. 

 The final step for all clusters for which this is feasible, was the determination of the kinematic outliers. We could only do that when a sample of at least 20 secure members,
identified independently, is available.
These firmly
identified cluster members were cross identified in the PM catalogue. 
From the selection of 20 or more bona fide cluster members, we estimated the statistical parameters of a 2D Gaussian model of the proper motions of the cluster in such a way that $\chi^2$ is close to 1. 
To this aim, we considered additional systematic uncertainties  
$\sigma_{\alpha,sys}$ and $\sigma_{\delta,sys}$ with typical values between 1 and 4 mas~yr$^{-1}$,
added in quadrature to the PM errors. The fitted proper-motion centroid of the cluster is ($\mu_{\alpha,\circ}$, $\mu_{\delta,\circ}$).  The final selection proceeded keeping a photometric candidate if 

$$ \sqrt{\frac{\left(\mu_{\alpha} - \mu_{\alpha,\circ}\right)^2}{\sigma^{2}_{\alpha} + \sigma^{2}_{\alpha,sys}} + 
\frac{\left(\mu_{\delta} - \mu_{\delta,\circ}\right)^2}{\sigma^{2}_{\delta} + \sigma^{2}_{\delta,sys}}} < 5 $$

\noindent or if there is no valid PM information; $\mu_{\alpha}$, $\mu_{\delta}$,
and the corresponding sigma are the PM in right ascension and declination and the associated errors. Finally, we should note that clusters with a number of bona fide kinematic members lower than 20, as well as those clusters without any previous kinematic membership study, were not analysed using this procedure and we had to rely only on the photometric data and isochrones to define possible targets. 

We wish to stress that we have a range of cases, from clusters for which it was
not possible to apply a kinematic membership  analysis and where the separation
of candidate members  depends entirely on the photometry, to  objects with a
careful kinematic description and separation between cluster  and field stars.
We worked with a wide variety of observational samples with different degrees of 
completeness and bias in the composition of the cluster and field mix. One of
the immediate aims of  the programme is the evaluation of the quality of candidate
member selection, using the RV and abundance data determined by the $Gaia$-ESO Survey and the astrometric information provided by the $Gaia$ mission (see Section~\ref{members}, the many $Gaia$-ESO papers, and Randich et al. 2021).

\subsection{The case of NGC\,6705}\label{NGC6705}

To better explain the selection process, we show an example of application
of this methodology to NGC\,6705, which was observed during the first six months of $Gaia$-ESO operations (May, June 2012). Literature parameters for this cluster are: age=250 Myr, $(m-M)_0=11.55$, E(B-V)=0.43 (see \citealt{sung}), and [Fe/H]=+0.10  \citep{gw}. These values are similar to those derived by \citet{cantatm11}, who used the data described below and the first $Gaia$-ESO results. Photometry in the $V$ and $B$ bands for about 22000 stars within a FoV with side of 22.5\arcmin\  were taken from \cite{koo}.   We selected candidate cluster stars, as shown in 
Fig.~\ref{ngc6705_fig1}(a), where red dots represent the  photometric
candidates along the cluster MS, as well as those stars lying on the
blue  straggler, binary sequence, and RC locations. 
We did that separately for the optical and near infrared (NIR)  CMDs. From the $BV$ selection, we matched
the optical photometry with 2MASS data. The infrared analysis was
based on the selection of the probable cluster members according to
their relative position on the CMDs. The selected
candidates are plotted in Fig. \ref{ngc6705_fig1}(b), where
red dots represent the  candidates after the NIR filtering.

Radial velocity measurements for stars in the field of NGC\,6705 were  secured
by \citet{kharchenko}, \citet{mermilliod}, and  \citet{frinchaboy}. In addition, HARPS data were taken from the ESO archive, originated by the programme `Search for Planets around Evolved Intermediate-Mass Stars' \citep{lovis}. 
Thus,  for our analysis we had a total of 38 bona fide cluster 
members (indicated in Fig.~\ref{ngc6705_fig1}(a)) with RV and PM measurements (the latter
come from the UCAC4 catalogue). This means that we could apply a selection based
on kinematic criteria, according to previous section.

\begin{figure}[h]
\begin{center}
\includegraphics[bb=40 135 520 600,clip,scale=0.45]{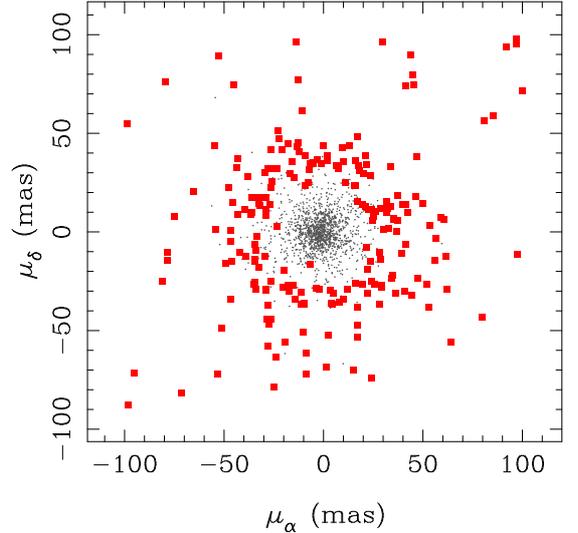}
\caption{Proper motion distribution for photometric candidate targets for NGC\ 6705 (grey symbols). Red filled squares represent kinematic outliers according to the procedure described in the text.}
\label{VPD}
\end{center}
\end{figure}

The determination of the cluster proper motion parameters was performed
following the procedure outlined above, yielding  -2.7 $\pm$ 0.7 mas~yr$^{-1}$
and -2.0 $\pm$ 0.5 mas~yr$^{-1}$  for the PM cluster centroid\footnote{The cluster average values obtained by \citet{cantat20} are $-1.560\pm0.160$ and $-4.144\pm0.161$ mas~yr$^{-1}$, respectively, based on more than 1000 stars down to $G=18$ measured by $Gaia$.}, with a systematic
error of 2.5 mas~yr$^{-1}$ and 1.3 mas~yr$^{-1}$,  respectively. To apply these
values, we matched the optical and infrared selected candidates with the PM
catalogue. The selection of the kinematic outliers and possible members of the
cluster was made according to these parameters down to V $\sim 17$. We found 226
stars whose proper motions are more than five sigma from the cluster centroid
(see Figs.~\ref{VPD} and \ref{ngc6705_fig1}(c)) and therefore they were excluded from the sample. Thus we selected a total of about 3400 candidate members, of which about 2250 do
not have kinematic information. 

For this cluster we had additional images in the {\em B, V}, and $I$ filters taken  with the Wide Field Imager at the Max-Planck-Gesellschaft/ESO 2.2m Telescope for the ESO Imaging Survey (ESO programme 163.O-0741(C), PI Renzini) in a field of 30$\times$30 arcmin$^2$. These data were reduced as described in \cite{cantatm11} who present $B$, $V$ and $I$ photometry  for more than 123000 stars down to V$\sim 22$. This catalogue was used for making the final selection and formed the basis for the preparation of the observing plan.

\begin{figure}[h]
\begin{center}
\includegraphics[scale=0.4]{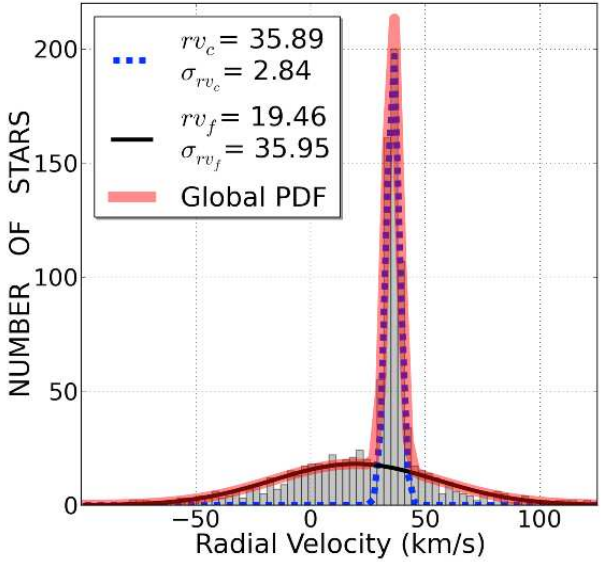} \caption{Distribution of the RVs of the 1028 stars observed
with GIRAFFE  in NGC~6705. The solid black line and the dashed blue
line represent the field stars and cluster members radial velocity
distributions. The probability density function (labelled PDF) of the total  sample is plotted
in red. As can be observed, a mean value of 35.89 km~s$^{-1}$ is found for the RV of the cluster
members, with a standard deviation of 2.84
km~s$^{-1}$.  For the field stars, the mean RV is 19.46
km~s$^{-1}$ and the standard deviation 35.95 km~s$^{-1}$.}
\label{ngc6705_mem}
\end{center}
\end{figure}

Of the sample described above, 1028 stars have been observed with GIRAFFE (860 with the HR15n
setup and 166 with the blue setups, see Table~\ref{tab1}) and 59  with UVES (ten with the setup U520, 49 with U580). 
These are shown in Fig.~\ref{ngc6705_fig1}, panel d, (red and blue dots for Giraffe and UVES, respectively)
along with the candidate members of the cluster (grey small dots). 
Post-observation membership  
is based here solely on RV values. An analysis of the radial velocity distribution reveals two Gaussian
populations whose statistical parameters are presented in Fig. \ref{ngc6705_mem}
and which lead to 536 stars classified as members of the cluster, representing 52\% of
the observed population. 

\begin{figure}
\begin{center}
\includegraphics[bb=60 180 480 570, clip,scale=0.55]{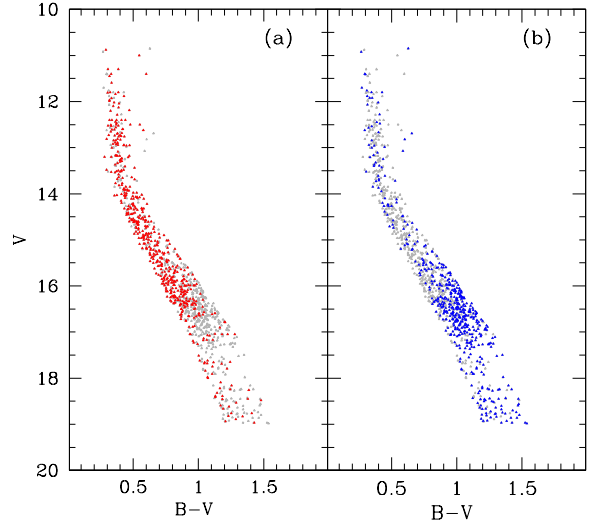}   \caption{(a) $V$, $B-V$ CMD for NGC 6705 showing the stars observed with GIRAFFE
in grey and  RV candidate cluster members in red. (b) As in previous panel, but with candidates non members indicated in blue. }
\label{ngc6705_GES}
\end{center}
\end{figure}

Figure~\ref{ngc6705_GES} shows our results, only for the stars observed with GIRAFFE (since the UVES fibres were generally allocated to stars known to be members); panels (a) and (b) display the CMD of stars found to be high probability member and non member, respectively.  For a more detailed treatment of this cluster see \cite{cantatm11}.

\section{The work-flow of the target selection and OB preparation} \label{workflow}

The clusters observed, their assigned  priority, and the driving science goals which motivated their study  are discussed elsewhere (Randich et al. 2021)\footnote{For the list of the 62 objects observed by GES, see Table~\ref{t:app} and Fig.~\ref{f:app}}. 
We only recall that the clusters have been selected taking into account the available information at the time of observations (essentially January 2012 to January 2018), especially regarding photometry and astrometry. In fact, we require the former to select target stars in the various evolutionary phases and the latter to allocate the FLAMES fibres and produce the so-called OBs, that is the suite of information needed to perform the observation.

We collected all relevant photometric data, including literature sources, private
data, and on-going surveys. A large effort was devoted to collecting all
existing  literature information on the proposed clusters. Their properties and
parameters are often rather  controversial. 
The stars were cross-identified to inter-compare and combine the photometric sets. We then identified the kind of stars to be observed
(see Sect.~\ref{whichtargets}). Unified criteria were used to determine instrumental setups and exposure times on
the basis of target spectral type (e.g. late-type dwarf versus early type dwarf), target
magnitude, fibre type (GIRAFFE or UVES), and scientific goals (e.g. RV vs. abundances determination). 

When possible, we performed a kinematic membership analysis, as described in
Sect.~\ref{methodology}. 
For the others, the $Gaia$-ESO survey is the first
source of kinematical information.
Furthermore, the number of observed stars  and for which we obtain at least 
RV data, largely exceeds all existing samples for all clusters.

During the target selection, care was taken to find all ESO archive observations
regarding  the clusters, especially those containing  spectroscopic data.  This helped to exclude
stars already observed, or to include a sub-sample of them for calibration and testing purposes. Furthermore, all the archive  observations compatible with the $Gaia$-ESO goals
(similar resolution spectra,  wavelength coverage,  etc) have been re-analysed
homogeneously, to increase the $Gaia$-ESO sample.

To avoid mis-centring of the targets in the fibres, we ensured that the coordinates of all targets, as well as those of
auxiliary stars used for guiding and alignments, were on a precise and homogeneous astrometric system. We used coordinates from 2MASS or from
other catalogues with better precision, registered to the 2MASS system.  All guide and fine-guidance stars were chosen to have possibly low proper motion (even if the observing software can compensate for that). At least 15 GIRAFFE and one UVES fibres were always assigned to sky
positions,  chosen so to follow as much as possible the spatial
distribution of science targets. For fields with strong and spatially variable
sky emission (e.g. some SFRs), more GIRAFFE sky fibres were  allocated,  to
improve the subtraction of sky features \citep[see e.g.][]{bonito20}.

Once the fibre allocation was done, we prepared binary FITS tables with information on
astrometry and  photometry for all observed stars (one per each OB), to be added
to the reduced  files that could be used for the scientific analysis and finally
made public through the ESO dedicated  archive. The use of FITS tables ensures
format uniformity and portability.

Tables~\ref{exptime1} and
\ref{exptime2} give the exposure times for representative cases of late-type
and early-type stars, respectively. All observations, as described by the OBs, were split into two or more equal
exposures for cosmic rays rejection and to allow for a short exposure with the
Th-Ar  wavelength calibration lamp on to obtain precise RVs (only five special fibres are illuminated in the {\em Simcal} observations). Short exposures
with the lamp on were interleaved with the science exposures for the following
setups: HR09b, HR14a, and HR15n. At the start of the Survey, for
HR03, HR04, HR05A, and HR6, the lamp was kept on during the science exposure.
As this led to contamination of some of the stellar data, the
practice was discontinued \citep{blomme21}.

The number of pointings and setups for each cluster, and thus the total observing
time, was chosen as to observe a large fraction, ideally $>80\%$, of
the candidate cluster members. However, this is difficult to generalise and in most cases impossible to reach. Sometimes photometry is available only on a limited area and does not cover the whole cluster extension. There are limitations due to the instrument itself, such as the FoV of 25\arcmin \ in diameter and the minimum fibre separation of 10.5\arcsec \ that makes pointing more difficult in the crowded central regions; we need to assign fibres to enough sky positions to ensure a reasonable background correction (in presence of variable extinction, such as in SFRs, their number and distribution must follow closely that of the stars); and we need to assign fibres  to guide stars, which take precedence over targets. Finally, we have only a limited amount of time assigned to each cluster, chosen to balance the sample of clusters and of stars observed in each of them (see Randich et al. 2021).

Given all the above limitations, a lower completeness fraction is acceptable, especially for distant and rich clusters, as well as for the outer parts of spatially extended clusters, when observing so many stars
becomes prohibitively time consuming. We can take as a sort of completeness parameter the ratio of possible targets to actually allocated fibres, which varies from about 20 to 80\%, with the lower values corresponding generally to the richer, more concentrated clusters (e.g. Trumpler~20, NGC~6705) and the higher to nearby, sparse objects (e.g. $\gamma$ Vel). The goal is to always observe a representative and significant sample of cluster stars and we deem to have reached it.

\begin{table}
\centering
\caption{Summary of exposure times for late type stars}
\begin{tabular}{llll}
\hline
Setup   & mag range      & exp. time   &S/N\\
        &                & (min)     &(goal)\\
\hline
\multicolumn{4}{c}{Old clusters$^a$} \\
UVES 580/520      & $V<13$         & 50          &$>60$\\
                  & $13.0<V<14.5$  & $3\times50$ &$>60$\\
                  & $15.5<V<16.5$  & $5\times50$ &$>60$\\
                  & $15.5<V<16.5$  & $7\times50$ &$>60$\\
GIRAFFE HR09b/15n & $13.0<V<15.0$  & 25  	&$>20$ \\
	          & $15.0<V<19.0$  & $2\times50$ &$>20$ \\
\hline
\multicolumn{4}{c}{SFRs and young clusters$^b$} \\
GIRAFFE HR15n     & $12.0<V<16.0$  & 20  	&$>15$ \\
	          & $16.0<V<19.0$  & 50  	&$>15$ \\
UVES 580/520      & $V<12.0$   & 20         &$>60$\\
                  & $V<13.5$       & 50  	 &$>60$\\
                  & $V<14.2$       & $2\times50$ &$>60$\\
	          & $V<15.2$       & $3\times50$ &$>60$\\
\hline
\end{tabular}
\begin{list}{}{}
\item[] $^a$ Exposure times driven by UVES observations.\\
$^b$ Exposure times driven by GIRAFFE observations.
\end{list}
\label{exptime1}
\end{table}

\begin{table*}
\begin{center}
\caption{Summary of exposure times (in min) for early-type stars.
}
\begin{tabular}{llrrrrrr}
\hline
Setup & mag range & \multicolumn{6}{c}{exposure time} \\\cline{3-8}
      &           & \multicolumn{2}{c}{original} & \multicolumn{2}{c}{revised (Dec 2013)} & \multicolumn{2}{c}{revised (Dec 2014)} \\
      &           & O,early-B & late-B,A & O,early-B & late-B,A & O,early-B & late-B,A \\
\hline
U520        & $8 < V  < 11 $ & \tablefootmark{a} & \tablefootmark{a} & \tablefootmark{a} & \tablefootmark{a} & \tablefootmark{a} & \tablefootmark{a}\\
HR03        & $10 < V < 13$ & ---       & ---      & ---       & ---       & 36        & --- \\
            & $11 < V < 14$ & 34        & ---      & 87        & ---       & 90        & --- \\
            & $12 < V < 15$ & 86        & 29       & {\it 110} & 74        & {\it 112} & 58  \\
            & $13 < V < 16$ & {\it 107} & 73       & ---       & {\it 93}  & ---       & {\it 75} \\
            & $13 < V < 17$ & ---       & {\it 94} & ---       & ---       & ---       & --- \\
HR04, HR05A & $10 < V < 13$ & ---       & ---      & ---       & ---       & 13        & --- \\
            & $11 < V < 14$ & 17        & ---      & 60        & ---       & 32        & --- \\
            & $12 < V < 15$ & 41        & 12       & {\it 75}  & 42        & {\it 40}  & 21  \\
            & $13 < V < 16$ & {\it 51}  & 31       & ---       & {\it 53}  & ---       & {\it 26} \\
            & $13 < V < 17$ & ---       & {\it 40} & ---       & ---       & ---       & --- \\
HR06        & $10 < V < 13$ & ---       & ---      & ---       & ---       & 9.3       & --- \\
            & $11 < V < 14$ & 14        & ---      & 46        & ---       & 23        & --- \\
            & $12 < V < 15$ & 35        & 10       & {\it 58}  & 33        & {\it 29}  & 15  \\
            & $13 < V < 16$ & {\it 43}  & 25       & ---       & {\it 41}  & ---       & {\it 19} \\
            & $13 < V < 17$ & ---       & {\it 32} & ---       & {\it 105} & ---       & {\it 49} \\
HR09B       & $10 < V < 13$ & ---       & ---      & ---       & ---       & 15        & --- \\
            & $11 < V < 14$ & 32        & ---      & 86        & ---       & 38        & --- \\
            & $12 < V < 15$ & 79        & 21       & {\it 108} & 56        & {\it 47}  & 25  \\
            & $13 < V < 16$ & {\it 99}  & 54       & ---       & {\it 70}  & ---       & {\it 32} \\
            & $13 < V < 17$ & ---       & {\it 69} & ---       & ---       & ---       & --- \\
HR14A       & $10 < V < 13$ & ---       & ---      & ---       & ---       & 5.6       & --- \\
            & $11 < V < 14$ & 14        & ---      & 19        & ---       & 14        & --- \\
            & $12 < V < 15$ & 34        & ---      & 46        & ---       & 35        & --- \\
            & $13 < V < 16$ & {\it 42}  & 22       & {\it 57}  & 30        & {\it 44}  & 23  \\
            & $13 < V < 17$ & ---       & {\it 28} & ---       & {\it 38}  & ---       & {\it 29} \\
\hline
\end{tabular}
\tablefoot{
\tablefoottext{a}{Exposure times are set by the corresponding exposure times for GIRAFFE. \\
The original exposure times were based on the GIRAFFE exposure time calculator. 
However, we experienced
problems attaining the required S/N and the integration
times were increased in December 2013. With the introduction of updated
values in the GIRAFFE exposure time calculator, these were again revised
in December 2014. The integration times aim for a S/N $>$ 100 for O,early B,
and $>$ 50 for late-B,A spectral types. For fainter stars, we cannot reach these 
values within a reasonable integration time, so the lower values of S/N $>$ 70 and 35, respectively, are aimed for; these integration times are set in italic. See \citet{blomme21} for details.}
}
\label{exptime2}
\end{center}
\end{table*}

\section{Actual fraction of member stars observed}\label{members}

To see how $Gaia$-ESO fared in observing cluster members, we may calculate the quotient between confirmed members and observed stars. In Table~\ref{fractio}, we report this ratio for a series of clusters already individually analysed, ordered by paper publication date. We take the numbers from the reference papers, indicated in the table, and use only studies and information related to GIRAFFE observations, since the UVES targets were pre-selected also on the basis of existing membership data. A caveat is that they relied on early releases and slightly different numbers might come out from analysis based on the final data release and the combination with $Gaia$ data (see e.g. \citealt{jackson20}). For intermediate-age and old clusters, the post-observation candidate members were selected based on their photometry and other literature parameters where possible (RV,  distance from the cluster centre and a metallicity similar to the bulk of the cluster population) as explained in previous section. For the young clusters, other indicators were generally preferred as more robust, such as lithium abundance (see the original papers for motivation and details).
The ratio generally ranges between $~15\%$ and $~60\%$, with older clusters having in general a better score.  We stress again that contaminants can be (and were) used for science and that in the close-by, young clusters the requirement of being unbiased means we had to accept the possibility of having a large number of contaminants.

We can check a posteriori how well our selection of candidate members fared in observing a representative sample  by asking how well we did in comparison with a completely independent selection based on $Gaia$ astrometry. Various groups studied OCs with $Gaia$ DR1 and especially DR2; one of the more complete samples of OCs and candidate members has been compiled by \citet{cantat18} and \citet{cantat20}. They provide tables with a membership probability for more than 2000 OCs. Table~\ref{fractioGaia} shows a few examples of a comparison for several clusters 
contained in a single FLAMES FoV (diameter 25 arcmin) for simplicity. 
We compare the number of stars observed and candidate members in our survey with the number of high probability members defined only on the basis of $Gaia$ astrometry. The columns give: the number of stars actually observed; the number within the same magnitude limit of  \citet{cantat18}, \citet{cantat20}, i.e. $G=18$; the number of stars in a region centred on each cluster with a radius of 12.5 arcmin in the $Gaia$-based catalogue; the numbers of stars with high astrometric membership probability (we choose here 0.7 and 0.9) in the same region; the number of members in each cluster in $Gaia$-ESO according to \citet[][see below]{jackson21}; and finally the ratios of $Gaia$-ESO to $Gaia$ candidate members. We see that we managed to observe a high fraction of possible members.

Deriving a membership fraction for all clusters is outside the scope of the present paper, as the methods and results depend on the scientific case one wishes to address. However, a more uniform analysis on a large sample of clusters was performed by \citet{jackson20,jackson21}. In the first paper, they used the internal DR5 of $Gaia$-ESO and $Gaia$ DR2  and studied the kinematical properties of 32 OCs. They determined the membership probability using a maximum likelihood analysis of the 3D velocity distributions (RV and tangential velocity computed on the basis of proper motions in right ascension and declination plus parallax) in each cluster, taking also into account variations in RV due to binarity.
In the second paper, the same kind of analysis was extended to 70 objects in iDR6, 63 open and seven globular clusters
\citep[][note that one of the Gaia-ESO targets, Loden~165, was not confirmed to be a real cluster.]{jackson21}. 

\begin{figure}
    \centering
    \includegraphics[scale=0.55]{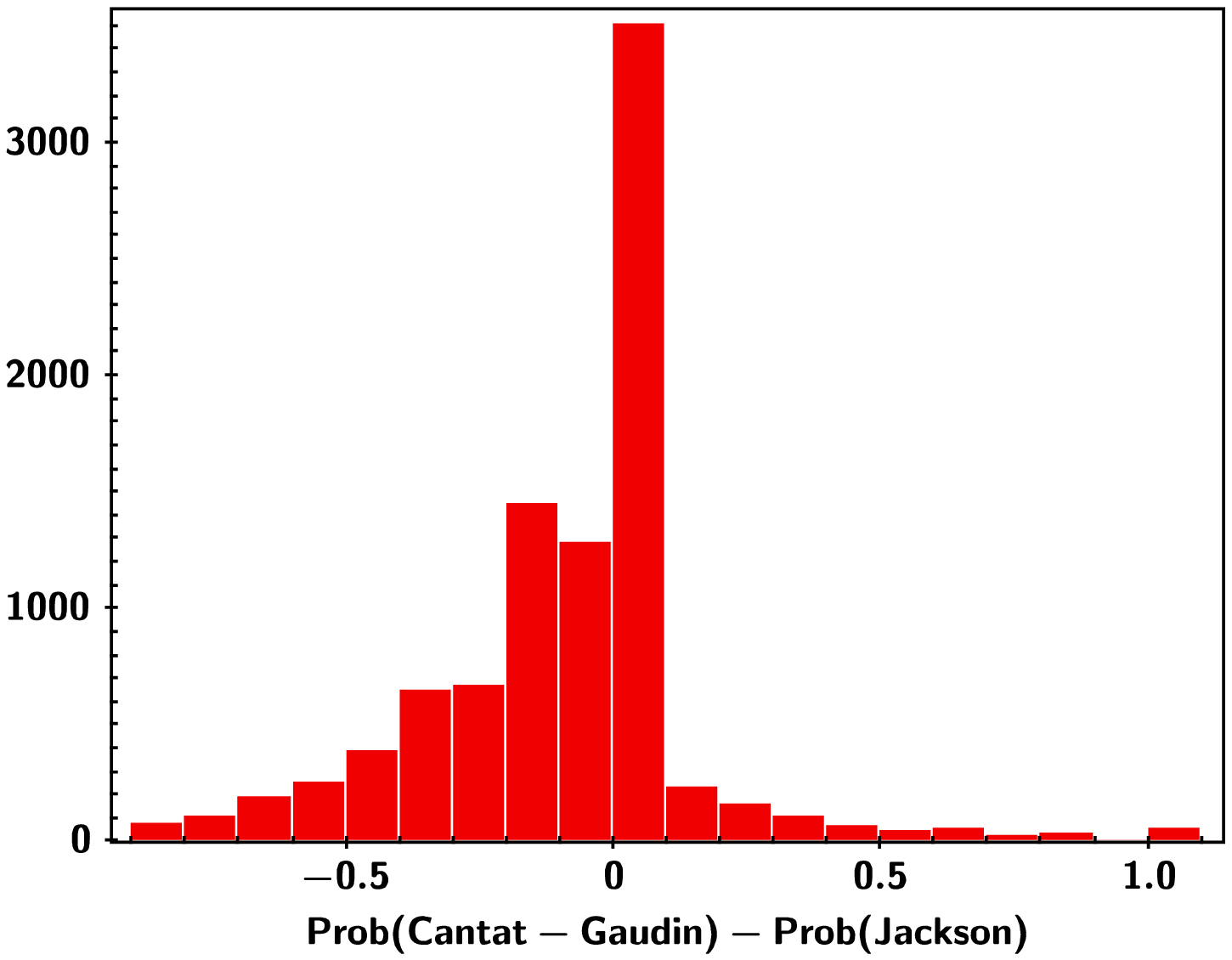}
    \caption{Histogram of the difference in membership probability measured by \citet{cantat18}, \citet{cantat20}, and \citet{jackson21}.}
    \label{deltaP}
\end{figure}

We can compare these results with membership probabilities derived on the basis of parallaxes and proper motions, as in \citet{cantat18}, \citet{cantat20}.  We have cross-matched their catalogue with the one in \citet[their Tables 3 and A2]{jackson21}. There are 9395 stars with valid membership probabilities measured in both catalogues\footnote{Note that $\gamma$~Vel is indicated as Pozzo~1 and $\lambda$~Ori as Collinder~69 in \citet{cantat18}.}; Fig.~\ref{deltaP} shows the difference  in membership probability. 
This difference has a peak near zero, with an average value of -0.12 (rms=0.27).
For 50\% of the stars the difference is within $\pm0.1$ and for more than 75\% is within $\pm0.3$ (4791 and 7306 stars over 9395 in total, respectively). 

More in depth comparisons with previous results and membership determination using other methods are presented in \citet{jackson20,jackson21}.
In particular, a discussion of the merit of using also RV in addition to proper motions (and parallax) to define membership can be found there. Very briefly, by adding the $Gaia$-ESO RVs, the fraction of false positives is efficiently reduced, meaning that a better separation of cluster members from the surrounding field can be reached. This is especially true for clusters at
larger distances.

A summary of the results is given in Table~\ref{t:app}, where the clusters are ordered by ascending age.
The fractions shown in the table should not be directly compared to the values in Table~\ref{fractio}, as only part of the observed targets could be used in the fit (see original paper for details).

\begin{table}
\caption{Fraction of GIRAFFE members in $Gaia$-ESO clusters already published. }
\begin{tabular}{llr}
\hline
Cluster                 & Fraction & Ref paper	       \\
\hline
Trumpler 20     	&  38\%   & \cite{donatitr20}        \\ 
$\gamma$ Vel       	&  17\%   & \cite{jeffriesGES}      \\ 
                	&  19\%   & \cite{prisinzano16}    \\ 
NGC 4815                &  29\%   & \cite{friel14}         \\ 
NGC 6705                &  52\%   & \cite{cantatm11} \\ 
Berkeley 81             &  16\%   & \cite{magrini15} \\
$\rho$ Oph        &  15\%   & \cite{rigliaco16}      \\ 
NGC 3293                &  24\%   & \cite{delgado016}       \\ 
Trumpler 23             &  40\%   & \cite{overbeek17}        \\ 
NGC 6802                &  52\%   & \cite{tang17}	       \\ 
Chamaleon I             &  13\%   & \cite{sacco17}	       \\ 
Carina (Tr14,Tr16,Cr232)&  36\%   & \cite{damiani17}         \\ 
IC 2602   		&   7\%   & \cite{bravi18}	       \\ 
IC 2391   		&  13\%   & \cite{bravi18}	       \\ 
IC 4665   		&  22\%   & \cite{bravi18}	       \\ 
NGC 2547                &  42\%   & \cite{bravi18}	       \\ 
Pismis 18$^a$               &  18\%   & \cite{hatzi19}  \\ 
NGC 2420                &  61\%   & \cite{semenova20}      \\ 
\hline
\end{tabular}
\tablefoot{
$^a$ For this cluster also $Gaia$ DR2 data were used to confirm membership, in addition to RV.}
\label{fractio}
\end{table}

\begin{table*}
\centering
\caption{Example of candidate members observed and available in a few clusters.}
\begin{tabular}{lrrrrrrcc}
\hline
cluster &N GES & N GES    & N C-T       & N C-T       & N C-T   & members    & GES/C-T  &GES/C-T \\
& observed   & $G\le18$ & d$\le12.5'$ & prob$\ge$0.7 &prob$\ge$0.9  & J+2021 &prob$\ge$0.7 & prob$\ge$0.9 \\
\hline
NGC~6709 & 730& 664 & 223 & 158 &  76 & 125 & 0.79 &1.64 \\
NGC~6705 &1066&1012 &2439 & 883 & 245 & 526 & 0.60 &2.15 \\
NGC~6802 & 197& 183 & 753 & 270 & 100 &  55 & 0.20 &0.55 \\
NGC~2355 & 208& 190 & 314 & 222 &  88 & 139 & 0.63 &1.58 \\
NGC~2158 & 616& 490 &1633 &1186 & 647 & 346 & 0.29 &0.53 \\
NGC~2420 & 562& 509 & 513 & 331 & 121 & 384 & 1.16 &3.17 \\
NGC~2243 & 710& 521 & 531 & 479 & 327 & 538 & 1.12 &1.65 \\
Berkeley~39  & 899& 675 & 562 & 508 & 382 & 507 & 1.00 &1.33 \\
Berkeley~36  & 739& 356 & 374 & 112 &  51 & 212 & 1.89 &4.16 \\
\hline
\end{tabular}
\tablefoot{
N GES is the number of stars in Gaia-ESO final data release; C-T stands for \citet{cantat18}, \citet{cantat20}; J+21 stands for \cite{jackson21}, their tables 4, 5; the last two columns give the fractions of stars assumed as member by J+21 and C-T (for the latter, membership probability larger than 0.7 and 0.9 were selected. }
\label{fractioGaia}
\end{table*}

\section{Summary}\label{summary}

An observational project such as the $Gaia$-ESO Survey, employing 340 VLT nights
over six years, required careful selection of targets to
maximise the effectiveness of the observational planning and ensure success in achieving its scientific objectives. We eventually observed 62 OCs and SFRs, see the list in the Appendix, Table~\ref{t:app}, and Randich et al. (2021) for details.

The $Gaia$-ESO consortium is organised in working groups
specialised in different tasks following a clear work-flow
\citep{gilmore12}, \citep{r&g13}. In this paper, we described how three of these
working groups prepared the actual observations. We focused, in particular,
on the crucial task of selecting  the target stars for each cluster which
would grant reaching the survey main goals, generating at the same time
the largest, more accurate, and more homogeneous spectroscopic data set on
star clusters ever achieved.

$Gaia$-ESO is an ESO legacy project: the survey, especially when combined
with the $Gaia$ results, has a strong impact on the astronomical
community. The raw data are fully available without proprietary time and $Gaia$-ESO is also
providing  advanced data products (reduced spectra, photometry,
stellar parameters, etc) that can be the basis for future studies, likely
with scientific goals different from the ones driving the survey. To fully
exploit this archive, future users must be aware of the complete process that generated the data. One of the key steps is the choice of the stars targeted in the many clusters.

Given the diversity of cutting-edge astronomical problems that can be
addressed through the analysis of star clusters and their very different
properties in terms of brightness and spectral types distribution, angular
size, surface density, richness, etc., the design of a single selection
procedure applicable to all clusters is an impossible task. However, we
believe that there are some common patterns that underlie all cases and
define the backbone of the procedure described in
Sects~\ref{emilio} and \ref{workflow}. In particular, we
stressed our decision to be as inclusive and unbiased as possible in our selection of targets, and
to discard only the clear outliers (kinematic or photometric).

The CMDs, both in optical and NIR ranges, formed the primary basis for the
selection of candidate targets, according to the fiducial evolutionary lines. 
In addition, we tried, as far as possible, to use the
available kinematic information to eliminate secure non-members. However,
this is conditioned by the physical characteristics of the
clusters and the existence of prior membership studies, making it
impossible to generalise to the entire sample. 
Only about 40\% of the
clusters observed included kinematic analysis for the selection
of targets.
As a good example of our protocol when kinematic selection was possible, we
described in some detail its application to the intermediate-age, rich
cluster NGC~6705.

The last step before observations consisted in the preparation of the OBs,
allocating the FLAMES fibres to the targets. The effectiveness and
completeness of the fibres' allocation process depend on many factors (area covered by the cluster, richness, concentration, and magnitude of the targets) but in all
cases we observed a significant fraction of all possible targets. 

The $Gaia$-ESO spectra are the first
kinematic data for many clusters in the survey.   They
represent an increase of one or two orders of magnitude in the  number  of cluster stars even for the clusters previously studied. The $Gaia$-ESO data for stellar clusters will have a strong legacy also for combination with future ground-based surveys, both spectroscopic (e.g. WEAVE, 4MOST) or photometric
(e.g. LSST at the Rubin telescope).

\acknowledgements{
We have made use of the WEBDA database,  originally developed by J-C.
Mermilliod, now operated at the Department of Theoretical Physics and
Astrophysics of the Masaryk University. AB thanks P. Montegriffo for  his
software CataPack. This research has made use of the SIMBAD database \citep{wenger}, operated at CDS, Strasbourg, France and of the VizieR catalogue access tool, CDS, Strasbourg, France (DOI: 10.26093/cds/vizier). The original description of the VizieR service was published in \citet{vizierori}. This research has made use of NASA's Astrophysics Data System. We made extensive use of TOPCAT (http://www.starlink.ac.uk/topcat/, \citealt{topcat}).
We benefited from discussions in
various $Gaia$-ESO workshops supported by the ESF (European Science Foundation)
through the GREAT ($Gaia$ Research for European Astronomy Training) Research
Network Program.
This work has made use of data from the European Space Agency (ESA) mission
{\it Gaia} (https://www.cosmos.esa.int/gaia), processed by the {\it Gaia}
Data Processing and Analysis Consortium (DPAC,
https://www.cosmos.esa.int/web/gaia/dpac/consortium). Funding for the DPAC
has been provided by national institutions, in particular the institutions
participating in the {\it Gaia} Multilateral Agreement. We acknowledge the support from INAF and Ministero
dell'Istruzione, dell'Universit\`a e della Ricerca (MIUR) in the form of the
grants `Premiale VLT 2012' and `The Chemical and Dynamical
Evolution of the Milky Way and Local Group Galaxies' (prot. 2010LY5N2T). E.J.A. acknowledges the financial support  by the
Spanish Ministerio de Educacion y Ciencia, through grant AYA2010-17631, and by
the Consejeria de Educacion y Ciencia de la Junta de Andalucia, through TIC101
and P08-TIC-4075. This work was partially supported by the $Gaia$ Research for
European Astronomy Training (GREAT-ITN) Marie Curie network, funded through the
European Union Seventh Framework Programme [FP7/2007-2013] under grant agreement
no 264895. This work was partly supported by the European Union FP7 programme
through ERC grant no 320360. This work was partly supported by the Leverhulme
Trust through grant RPG-2012-541. T.B. was funded by grant No. 621-2009-3911 and grant No. 2018-0485 from
The Swedish Research Council. E.J.A. also acknowledges ESF GREAT grant No 6901.
R.S. acknowledges support from the National Science Centre, Poland (2014/15/B/ST9/03981). U.H. acknowledges support from the Swedish National Space Agency (SNSA/Rymdstyrelsen). F.J.E. acknowledges financial support from the Spanish MINECO/FEDER
through the grant AYA2017-84089 and MDM-2017-0737 at Centro de
Astrobiologia (CSIC-INTA), Unidad de Excelencia Maria de Maeztu, and
from the European Union's Horizon 2020 research and innovation programme
under Grant Agreement no. 824064 through the ESCAPE - The European
Science Cluster of Astronomy \& Particle Physics ESFRI Research
Infrastructures project.}

\clearpage 
\begin{appendix}\label{app}
\section{Open clusters observed by GES}

\begin{figure}
\centering
\includegraphics[scale=0.5]{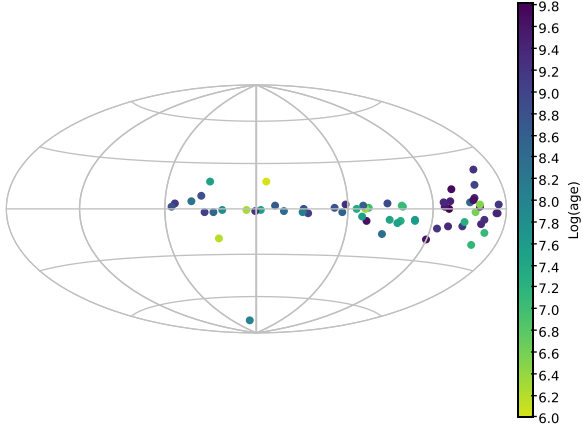}
\caption{Plot of all objects observed by GES in Galactic coordinates. The clusters are colour-coded by their age.}
\label{f:app}
\end{figure}

Figure~\ref{f:app} shows the Galactic positions of all OCs  observed by $Gaia$-ESO.  Table~\ref{t:app} gives the complete list of names in ascending age order and information on them. Three clusters are separated from the main list: M67 and NGC~6253 were observed as calibrators and their selection of targets is not the same of the main sample; Loden~165 has not been confirmed as a genuine open cluster (its coordinates and age come from \citealt{carraro01}).

Equatorial coordinates and ages come from \citet{tristan20} as in Randich et al. (2021), except for the ages of
NGC~2244, NGC~6530 \citep{bell13}; NGC~2264 \citep{venuti18};
$\gamma$~Vel \citep{franciosini21}; Trumpler~14 \citep{damiani17}; $\rho$~Oph \citep{grasser21}; and Chamaleon I \citep{galli21}. The column Class indicates the class, as defined in Sects~2 and 3: 1a is for young OCs, with no or few massive stars, 1b is for massive-star young clusters, and 2 is for older clusters. Metallicity based on GES published papers is indicated in column [Fe/H], taken from Randich et al. (2021); for NGC~3293, NGC~3766, and NGC~6649 values come from unpublished iDR6 results. 
Information on membership comes from \citet[][Tables 1, 2, 4, and 5]{jackson21} and is available for almost all clusters.
We note that N$_{obs}$ is the number of targets observed in each cluster with UVES and GIRAFFE HR15n (the main sample analysed by \citealt{jackson21});
N$_{comp}$ is the number of targets with a full set of the required data (2MASS $K_s$, $Gaia$ G, $Gaia$-ESO T$_{eff}$, and S/N$>$5, see their Sect. 2);
N$_{fit}$ is the number of targets actually fitted in  the membership analysis. 
Fractio M is the fraction of the N$_{fit}$ targets analysed and with a valid membership probability (i.e. not $-1$ in their Table~3) that are expected to be cluster members. In a few cases, these four values are absent in \cite{jackson21}.

\begin{table*}
\centering
\caption{Information on the $Gaia$-ESO observed clusters. } 
\begin{tabular}{lrrlrrrrrc}
\hline
Cluster & RA & Dec & age    &Class & [Fe/H]     & N$_{obs}$ & N$_{comp}$ & N$_{fit}$ &Fractio M \\
        &(J2000) &(J2000) & (Gyr)& & &	      &  	  &	      & 	 \\
\hline       
\object{$\rho$ Oph}           	   &16:24:00.00 &-23:48:00.0 & 0.001 &1a & 0.03 & 311 &301   & 72  &0.63 \\
\object{Chamaeleon I}         	   &16:24:00.00 &-23:48:00.0 & 0.001 &1a &-0.03 & 708 &687   &170  &0.55 \\
\object{NGC 6530}             	   &18:04:21.60 &-24:19:48.0 & 0.002 &1b &-0.02 &1972 &1907  &1501 &0.34 \\
\object{Trumpler 14}          	   &10:43:56.64 &-59:33:10.8 & 0.003 &1b &-0.01 &1111 &1069  &741  &0.53 \\
\object{NGC 2264}             	   &06:40:52.08 &+09:52:37.2 & 0.003 &1b &-0.10 &1876 &1819  &1408 &0.38 \\
\object{NGC 2244}             	   &06:32:10.80 &+04:54:50.4 & 0.004 &1b &-0.04 & 432 & 427  & 375 &0.35 \\
\object{NGC 3293}             	   &10:35:52.80 &-58:13:51.6 & 0.010 &1b &-0.08 &     &      &     &	   \\
\object{ASCC 50}/\object{Alessi 43}    	     &08:50:31.44 &-41:44:16.8 & 0.011 &1a & 0.02 &1224 &1192  & 501 &0.41 \\
\object{Collinder 69}/\object{$\lambda$ Ori} &05:35:10.08 &+09:48:46.8 & 0.013 &1a &-0.09 & 608 &588   & 344 &0.60 \\
\object{25 Ori}               	   &05:24:47.52 &+01:39:18.0 & 0.013 &1a & 0.00 & 294 & 284  & 256 &0.68 \\ 
\object{Collinder 197}        	   &08:44:48.48 &-41:16:48.0 & 0.014 &1a & 0.03 & 409 & 395  & 334 &0.37 \\
\object{NGC 2232}             	   &06:27:33.12 &-04:44:56.4 & 0.018 &1a &-0.03 &1761 &1734  & 697 &0.13 \\
\object{$\gamma$ Vel}         	   &08:09:29.76 &-47:20:06.0 & 0.020 &1a &-0.02 &1262 &1242  & 497 &0.45 \\
\object{NGC 3766}            	   &11:36:14.64 &-61:36:57.6 & 0.023 &1b &-0.12 &     &      &     &	 \\
\object{IC 2391}              	   &08:41:10.08 &-52:59:27.6 & 0.029 &1a &-0.06 & 434 &426   &78   &0.61 \\
\object{NGC 2547 }            	   &08:10:06.00 &-49:11:52.8 & 0.032 &1a &-0.03 & 477 &472   &269  &0.62 \\
\object{IC 4665 }             	   &17:46:12.96 &+05:36:54.0 & 0.033 &1a & 0.01 & 567 &562   & 298 &0.11 \\
\object{NGC 6405}             	   &17:40:16.56 &-32:14:31.2 & 0.035 &1a &-0.02 & 659 & 654  &373  &0.19 \\
\object{NGC 2451A}            	   &07:42:56.64 &-38:15:50.4 & 0.035 &1a &-0.08 &1656 &1637  &352  &0.13 \\
\object{IC 2602}              	   &10:42:27.12 &-64:25:33.6 & 0.036 &1a &-0.06 &1840 &1817  &117  &0.53 \\
\object{NGC 2451B}            	   &07:44:30.72 &-37:57:14.4 & 0.041 &1a &-0.02 &1656 &1635  &425  &0.16 \\
\object{NGC 6649}             	   &18:33:26.16 &-10:23:56.4 & 0.071 &1b &-0.08 & 122 & 121  &116  &0.62 \\
\object{Blanco 1}             	   &00:03:24.72 &-29:57:28.8 & 0.105 &2  &-0.03 & 463 &446   &314  &0.43 \\
\object{NGC 6067}             	   &16:13:11.76 &-54:13:37.2 & 0.126 &2  & 0.03 & 532 &531   &512  &0.39 \\
\object{NGC 6709}             	   &18:51:20.64 &+10:20:02.4 & 0.191 &2  &-0.02 & 684 & 681  &551  &0.15 \\
\object{NGC 2516}             	   &07:58:06.48 &-60:48:00.0 & 0.240 &2  &-0.04 & 759 &745   &641  &0.75 \\
\object{NGC 6259}             	   &17:00:46.80 &-44:40:40.8 & 0.269 &2  & 0.18 & 438 &423   &391  &0.38 \\
\object{Berkeley 30}          	   &06:57:45.12 &+03:13:44.4 & 0.295 &2  &-0.13 & 226 & 224  &216  &0.34 \\
\object{NGC 6705}             	   &18:51:03.84 &-06:16:19.2 & 0.309 &2  & 0.03 &1066 &1042  &977  &0.59 \\
\object{NGC 4815}             	   &12:57:59.76 &-64:57:36.0 & 0.372 &2  & 0.08 & 126 &126   &112  &0.50 \\
\object{NGC 3532}             	   &11:05:40.08 &-58:42:25.2 & 0.398 &2  &-0.01 & 966 & 952  &687  &0.73 \\
\object{NGC 6281}             	   &17:04:42.96 &-37:56:52.8 & 0.513 &2  &-0.04 & 251 & 249  & 63  &0.44 \\
\object{Pismis 18}            	   &13:36:54.48 &-62:05:27.6 & 0.575 &2  & 0.14 & 101 &101   & 90  &0.31 \\
\object{NGC 6802}             	   &19:30:36.24 &+20:15:43.2 & 0.661 &2  & 0.14 & 103 &103   & 98  &0.56 \\
\object{NGC 6633}             	   &18:27:22.80 &+06:36:54.0 & 0.692 &2  &-0.03 &1595 & 363  &119  &0.21 \\
\object{Trumpler 23}          	   &16:00:52.32 &-53:32:20.4 & 0.708 &2  & 0.20 &  89 &89    & 83  &0.47 \\
\object{Pismis 15}            	   &09:34:44.16 &-48:02:24.0 & 0.871 &2  & 0.02 & 235 & 235  &224  &0.19 \\
\object{NGC 2355}             	   &07:16:59.28 &+13:46:19.2 & 1.000 &2  &-0.13 & 208 & 208  &204  &0.69 \\
\object{Berkeley 81}          	   &19:01:40.56 &-00:27:14.4 & 1.148 &2  & 0.22 & 203 &203   &171  &0.34 \\
\object{NGC 6005}             	   &15:55:49.20 &-57:26:20.4 & 1.259 &2  & 0.22 & 355 &353   &325  &0.19 \\
\object{Berkeley 73}          	   &06:22:04.80 &-06:19:15.6 & 1.413 &2  &-0.26 &  76 & 75   & 70  &0.66 \\
\object{Berkeley 44}          	   &19:17:15.12 &+19:33:00.0 & 1.445 &2  & 0.22 &  93 &92    &83   &0.51 \\
\object{NGC 2158}             	   &06:07:26.88 &+24:05:56.4 & 1.549 &2  &-0.15 & 616 & 598  & 571 &0.67 \\
\object{Ruprecht 134}         	   &17:52:44.16 &-29:32:13.2 & 1.660 &2  & 0.27 & 680 &665   &602  &0.18 \\
\object{NGC 2420}             	   &06:49:00.48 &-23:59:56.4 & 1.698 &2  &-0.15 & 562 &557   &520  &0.75 \\
\object{Berkeley 75}          	   &07:38:24.48 &+21:34:30.0 & 1.738 &2  &-0.34 &  75 & 74   & 64  &0.71 \\
\object{NGC 2141}             	   &06:02:56.16 &+10:27:03.6 & 1.862 &2  &-0.04 & 853 & 846  &801  &0.76 \\
\object{Trumpler 20}          	   &12:39:31.68 &-60:38:13.2 & 1.862 &2  & 0.13 & 552 &545   &490  &0.38 \\
\object{Berkeley 21}          	   &05:51:43.20 &+21:48:43.2 & 2.138 &2  &-0.21 & 744 & 738  & 574 &0.34 \\
\object{NGC 2425}             	   &07:38:18.48 &-14:53:06.0 & 2.399 &2  &-0.13 & 528 & 525  &481  &0.31 \\ 
\object{Berkeley 22}          	   &05:58:28.32 &+07:45:46.8 & 2.455 &2  &-0.26 & 395 & 395  & 352 &0.52 \\
\object{Berkeley 25}         	   &06:41:16.08 &-16:29:13.2 & 2.455 &2  &-0.25 &     &	     &     &    \\
\object{Czernik 24}           	   &05:55:23.52 &+20:52:33.6 & 2.692 &2  &-0.11 & 346 & 343  &302  &0.27 \\
\object{Berkeley 31}          	   &06:57:37.44 &+08:17:06.0 & 2.818 &2  &-0.29 & 616 &614   &499  &0.34 \\
\object{Czernik 30}           	   &07:31:11.04 &-09:56:42.0 & 2.884 &2  &-0.31 & 226 & 226  & 193 &0.38 \\
\object{Haffner 10}           	   &07:28:37.44 &-15:21:50.4 & 3.802 &2  &-0.10 & 460 & 457  &428  &0.62 \\
\object{Trumpler 5}           	   &06:36:30.24 &+09:27:54.0 & 4.266 &2  &-0.35 &1138 &1132  &1098 &0.76 \\
\object{NGC 2243}             	   &06:29:34.80 &-31:16:55.2 & 4.365 &2  &-0.45 & 703 &701   &614  &0.88 \\
\object{ESO 92-05}            	   &10:03:12.24 &-64:45:18.0 & 4.467 &2  &-0.29 & 212 & 210  & 114 &0.80 \\
\object{Berkeley 32}          	   &06:58:07.20 &+06:25:58.8 & 4.898 &2  &-0.31 & 389 & 385  &348  &0.70 \\
\object{Berkeley 39}          	   &07:46:48.48 &-04:39:54.0 & 5.623 &2  &-0.14 & 899 & 897  &832  &0.62 \\      
\object{Berkeley 36}          	   &07:16:25.20 &-13:11:45.6 & 6.761 &2  &-0.15 & 739 & 737  &672  &0.34 \\
\hline
\object{Messier 67}           	   &08:51:23.04 &+11:48:50.4 & 3.981 &-  &-0.02 & 131 & 131  & 130 &0.92 \\
\object{NGC 6253}             	   &16:59:06.72 &-52:42:43.2 & 3.246 &-  & 0.33 & 294 & 235  & 227 &0.72 \\
\hline
\object{Loden 165}            	   &10:35:56	&-58:44:03   & 3     &2  &      & 388 & 387  &333  &    \\	     
\hline						       
\end{tabular}					      
\label{t:app}				       
\end{table*}

\end{appendix}


\begin{thebibliography}{}

\bibitem[Baume et al.(2003)]{baume} Baume, G., V{\'a}zquez, R.~A., Carraro, G., \& Feinstein, A.\ 2003, \aap, 402, 549 

\bibitem[Blomme et al. (2021)]{blomme21} Blomme, R., Daflon, S.,  Gebran. M., et al, \aap, submitted

\bibitem[Bell et al.(2013)]{bell13} Bell, C.~P.~M., Naylor, T., Mayne, N.~J., et al.\ 2013, \mnras, 434, 806. doi:10.1093/mnras/stt1075

\bibitem[Bodenheimer(1965)]{bodenheimer} Bodenheimer, P.\ 1965, 
\apj, 142, 451 

\bibitem[Bonito et al.(2020)]{bonito20} Bonito, R., Prisinzano, L., Venuti, L., et al.\ 2020, \aap, 642, A56. doi:10.1051/0004-6361/202037942

\bibitem[Bravi et al.(2018)]{bravi18} Bravi, L., Zari, E., Sacco, G.~G., et al.\ 2018, \aap, 615, A37. doi:10.1051/0004-6361/201832645

\bibitem[Bressan et al.(2012)]{parsec} Bressan, A., Marigo,  P., Girardi, L., et al.\ 2012, \mnras, 427, 127 

\bibitem[Cantat-Gaudin et al.(2014)]{cantatm11}
Cantat-Gaudin, T., Vallenari, A., Zaggia, S., et al.\ 2014, \aap, 569, A17

\bibitem[Cantat-Gaudin et al.(2018)]{cantat18} Cantat-Gaudin, T., Jordi, C., Vallenari, A., et al.\ 2018, \aap, 618, A93

\bibitem[Cantat-Gaudin \& Anders(2020)]{cantat20} Cantat-Gaudin, T. \& Anders, F.\ 2020, \aap, 633, A99

\bibitem[Cantat-Gaudin et al.(2020)]{tristan20} Cantat-Gaudin, T., Anders, F., Castro-Ginard, A., et al.\ 2020, \aap, 640, A1. doi:10.1051/0004-6361/202038192

\bibitem[Carraro et al.(2001)]{carraro01} Carraro, G., Patat, F., \& Baumgardt, H.\ 2001, \aap, 371, 107. doi:10.1051/0004-6361:20010307

\bibitem[Carraro et al.(2010)]{carraro} Carraro, G., Costa, E., 
\& Ahumada, J.~A.\ 2010, \aj, 140, 954 

\bibitem[Castro-Ginard et al.(2020)]{castro20} Castro-Ginard, A., Jordi, C., Luri, X., et al.\ 2020, \aap, 635, A45

\bibitem[Daflon \& Cunha(2004)]{daflon} Daflon, S., \& Cunha, K.\ 2004, \apj, 617, 1115 

\bibitem[Damiani et al.(2017)]{damiani17} Damiani, F., Klutsch, A., Jeffries, R.~D., et al.\ 2017, \aap, 603, A81. doi:10.1051/0004-6361/201629020

\bibitem[Delgado et al.(2007)]{delgado} Delgado, A.~J., Alfaro, E.~J., \& Yun, J.~L.\ 2007, \aap, 467, 1397

\bibitem[Delgado et al.(2016)]{delgado016} Delgado, A.~J., Sampedro, L., Alfaro, E.~J., et al.\ 2016, \mnras, 460, 3305. doi:10.1093/mnras/stw1217

\bibitem[Damiani et al.(2017)]{damiani17} Damiani, F., Klutsch, A., Jeffries, R.~D., et al.\ 2017, \aap, 603, A81. doi:10.1051/0004-6361/20162902

\bibitem[Donati et al.(2014a)]{donati3oc} Donati, P., Beccari, G., Bragaglia, A., et al.\ 2014a, \mnras, 437, 1241. doi:10.1093/mnras/stt1944

\bibitem[Donati et al.(2014b)]{donatitr20} Donati, P., Cantat Gaudin, T.,
Bragaglia, A., et al. 2014b, \aap, 561, A94

\bibitem[Donor et al.(2020)]{donor20} Donor, J., Frinchaboy, P.~M., Cunha, K., et al.\ 2020, \aj, 159, 199

\bibitem[Drew et al.(2005)]{iphas} Drew, J.~E., Greimel, R., 
Irwin, M.~J., et al.\ 2005, \mnras, 362, 753 

\bibitem[Drew et al.(2014)]{vphas} Drew, J.~E., 
Gonzalez-Solares, E., Greimel, R., et al.\ 2014, \mnras, 440, 2036 

\bibitem[Favata \& Micela(2003)]{favata_micela}  Favata, F., \& Micela, G.\ 2003, \ssr, 108, 577

\bibitem[Feigelson et al.(2007)]{feigelson} Feigelson, E.,  Townsley, L.,
G{\"u}del, M.,  \& Stassun, K.\ 2007, Protostars and Planets V, 313

\bibitem[Franciosini et al.(2021)]{franciosini21}Franciosini, E., Tognelli, E., Degl’Innocenti, S., et al. 2021, \aap, in press, arXiv:2111.11196

\bibitem[Friel(1995)]{friel95} Friel, E.~D.\ 1995, \araa, 33, 381 

\bibitem[Friel et al.(2014)]{friel14} Friel, E.~D., Donati, P., Bragaglia, A., et al.\ 2014, \aap, 563, A117. doi:10.1051/0004-6361/201323215

\bibitem[Frinchaboy \& Majewski(2008)]{frinchaboy} Frinchaboy, P.~M., \& Majewski, S.~R.\ 2008, \aj, 136, 118 

\bibitem[Gaia Collaboration et al.(2016a)]{prusti} Gaia Collaboration, Prusti, T., de Bruijne, J.~H.~J., et al.\ 2016a, \aap, 595, A1. doi:10.1051/0004-6361/201629272

\bibitem[Gaia Collaboration et al.(2016b)]{GDR1a} Gaia Collaboration, Brown, A.~G.~A., Vallenari, A., et al.\ 2016b, \aap, 595, A2

\bibitem[Gaia Collaboration et al.(2017)]{GDR1b} Gaia Collaboration, van Leeuwen, F., Vallenari, A., et al.\ 2017, \aap, 601, A19

\bibitem[Gaia Collaboration et al.(2018)]{GDR2b} Gaia Collaboration, Babusiaux, C., van Leeuwen, F., et al.\ 2018, \aap, 616, A10

\bibitem[Gaia Collaboration et al.(2018)]{GDR2a} Gaia Collaboration, Brown, A.~G.~A., Vallenari, A., et al.\ 2018, \aap, 616, A1

\bibitem[Gaia Collaboration et al.(2022)]{eDR3} Gaia Collaboration, Brown, A.~G.~A., Vallenari, A., et al.\ 2021, \aap, 649, A21

\bibitem[Galli et al.(2021)]{galli21} Galli, P.~A.~B., Bouy, H., Olivares, J., et al.\ 2021, \aap, 646, A46. doi:10.1051/0004-6361/202039395

\bibitem[Gieles et al.(2006)]{gieles06} Gieles, M., Portegies Zwart, S. F., Baumgardt, H., et al. 2006, 
\mnras, 371, 793

\bibitem[Gilmore et al.(2012)]{gilmore12} Gilmore, G., Randich, 
S., Asplund, M., et al.\ 2012, The Messenger, 147, 25 

\bibitem[Gilmore et al.(2021)]{gilmoreinprep} Gilmore, G., Randich, S., et al., in preparation

\bibitem[Girard et al.(2011)]{girard} Girard, T.~M., van 
Altena, W.~F., Zacharias, N., et al.\ 2011, \aj, 142, 15 

\bibitem[Gonzalez \& Wallerstein(2000)]{gw} Gonzalez, G., \& Wallerstein, G.\ 2000, \pasp, 112, 1081 

\bibitem[Grasser et al.(2021)]{grasser21} Grasser, N., Ratzenb{\"o}ck, S., Alves, J., et al.\ 2021, \aap, 652, A2. doi:10.1051/0004-6361/202140438

\bibitem[Groot et al.(2009)]{groot} Groot, P.~J., Verbeek,  K., Greimel, R., et al.\ 2009, \mnras, 399, 323 

\bibitem[Hatzidimitriou et al.(2019)]{hatzi19} Hatzidimitriou, D., Held, E.~V., Tognelli, E., et al.\ 2019, \aap, 626, A90. doi:10.1051/0004-6361/201834636

\bibitem[Hayes \& Friel(2013)]{hayes_friel} Hayes, C., Friel, E.D., 2013, \aj, 147, 69

\bibitem[Jackson et al.(2015)]{jackson15} Jackson, R.~J., Jeffries, R.~D., Lewis, J., et al.\ 2015, \aap, 580, A75. doi:10.1051/0004-6361/201526248

\bibitem[Jackson et al.(2020)]{jackson20} Jackson, R.~J., Jeffries, R.~D., Wright, N.~J., et al.\ 2020, \mnras, 496, 4791 

\bibitem[Jackson et al.(2021)]{jackson21}  Jackson, R.~J., Jeffries, R.~D., Wright, N.~J., et al.\ 2021, \mnras. doi:10.1093/mnras/stab3032

\bibitem[Jadhav et al.(2021)]{jadhav21} Jadhav, V.~V., Pennock, C.~M., Subramaniam, A., et al.\ 2021, \mnras, 503, 236. doi:10.1093/mnras/stab213
\bibitem[Janes \& Phelps(1994)]{janes94} Janes, K.~A., \& Phelps, R.~L.\ 1994, \aj, 108, 1773 

\bibitem[Jeffries et al.(2009)]{jeffries} Jeffries, R.~D., 
Naylor, T., Walter, F.~M., Pozzo, M.~P., 
\& Devey, C.~R.\ 2009, \mnras, 393, 538 

\bibitem[Jeffries et al.(2014)]{jeffriesGES} Jeffries, R.~D., Jackson, R.~J., Cottaar, M., et al.\ 2014, \aap, 563, AA94 

\bibitem[Keller et al.(2007)]{skymapper} Keller, S.~C., Schmidt, 
B.~P., Bessell, M.~S., et al.\ 2007, \pasa, 24, 1 

\bibitem[Kharchenko et al.(2007)]{kharchenko} Kharchenko, N.~V., 
Scholz, R.-D., Piskunov, A.~E., R{\"o}ser, S., 
\& Schilbach, E.\ 2007, Astronomische Nachrichten, 328, 889 

\bibitem[Koo et al.(2007)]{koo} Koo, J.-R., Kim, S.-L., Rey, S.-C., et al.\ 2007, \pasp, 119, 1233 

\bibitem[Lada \& Lada(2003)]{lada} Lada, C.~J., \& Lada, E.~A.\ 2003, \araa, 41, 57 

\bibitem[Lindegren et al.(2018)]{GDR2d} Lindegren, L., et al.\ 2018, \aap, 616, A2

\bibitem[Liu \& Pang(2019)]{lp19} Liu, L. \& Pang, X.\ 2019, \apjs, 245, 32

\bibitem[Lovis \& Mayor(2007)]{lovis} Lovis, C., \& Mayor, M.\ 2007, \aap, 472, 657 

\bibitem[Magrini et al.(2015)]{magrini15} Magrini, L., Randich, S., Donati, P., et al.\ 2015, \aap, 580, A85. doi:10.1051/0004-6361/201526305

\bibitem[Magrini et al.(2021)]{magrini21} Magrini, L., Lagarde, N., Charbonnel, C., et al.\ 2021, \aap, 651, A84. doi:10.1051/0004-6361/202140935

\bibitem[McMahon(2012)]{vhs} McMahon, R. 2012, Science from the Next Generation
Imaging and Spectroscopic Surveys	

\bibitem[Mermilliod et  al.(2008)]{mermilliod} Mermilliod, J.~C., Mayor, M., \& Udry, S.\ 2008, \aap, 485, 303 

\bibitem[Mignard(2005)]{mignard} Mignard, F.\ 2005, Astrometry 
in the Age of the Next Generation of Large Telescopes, 338, 15 

\bibitem[Ochsenbein et al.(2000)]{vizierori} Ochsenbein F., Bauer P., Marcout J., 2000, \aaps, 143, 23

\bibitem[Overbeek et al.(2017)]{overbeek17} Overbeek, J.~C., Friel, E.~D., Donati, P., et al.\ 2017, \aap, 598, A68. doi:10.1051/0004-6361/201629345

\bibitem[Pancino et al.(2017)]{pancino17} Pancino, E., Lardo, C., Altavilla, G., et al.\ 2017, \aap, 598, A5. doi:10.1051/0004-6361/201629450

\bibitem[Pasquini et al.(2002)]{flames} 
Pasquini, L. et al. 2002, The Messenger 110, 1

\bibitem[Prisinzano et al.(2016)]{prisinzano16} Prisinzano, L., Damiani, F., Micela, G., et al.\ 2016, \aap, 589, A70. doi:10.1051/0004-6361/201527875

\bibitem[Prisinzano et al.(2019)]{prisinzano19} Prisinzano, L., Damiani, F., Kalari, V., et al.\ 2019, \aap, 623, A159. doi:10.1051/0004-6361/201834870

\bibitem[Randich et al.(2013)]{r&g13} Randich, S., Gilmore, 
G., \& Gaia-ESO Consortium 2013, The Messenger, 154, 47 

\bibitem[Randich et al.(2018)]{randichtgas} Randich, S., Tognelli, E., Jackson, R., et al.\ 2018, \aap, 612, A99

\bibitem[Randich et al.(2021)]{randichinprep} Randich, S., Gilmore, G., et al., in preparation

\bibitem[Rigliaco et al.(2016)]{rigliaco16} Rigliaco, E., Wilking, B., Meyer, M.~R., et al.\ 2016, \aap, 588, A123. doi:10.1051/0004-6361/201527253

\bibitem[R{\"o}ser et al.(2008)]{roeser} R{\"o}ser, S., Schilbach, E., Schwan, H., et al.\ 2008, \aap, 
488, 401 

\bibitem[Romano et al.(2021)]{romano21} Romano, D., Magrini, L., Randich, S., et al.\ 2021, \aap, 653, A72. doi:10.1051/0004-6361/202141340

\bibitem[Sacco et al.(2017)]{sacco17} Sacco, G.~G., Spina, L., Randich, S., et al.\ 2017, \aap, 601, A97. doi:10.1051/0004-6361/201629698

\bibitem[Sartoretti et al.(2018)]{GDR2c} Sartoretti, P., Katz, D., Cropper, M., et al.\ 2018, \aap, 616, A6

\bibitem[Semenova et al.(2020)]{semenova20} Semenova, E., Bergemann, M., Deal, M., et al.\ 2020, \aap, 643, A164. doi:10.1051/0004-6361/202038833

\bibitem[Siess et  al.(2000)]{siess} Siess, L., Dufour, E., \& Forestini, M.\ 2000, \aap, 358, 593 

\bibitem[Sim et al.(2019)]{sim19} Sim, G., Lee, S.~H., Ann, H.~B., et al.\ 2019, Journal of Korean Astronomical Society, 52, 145

\bibitem[Spina et al.(2014)]{spina} Spina, L., Randich, S., Palla, F., et al.\ 2014, \aap, 567, AA55 

\bibitem[Skrutskie et al.(2006)]{2mass} Skrutskie, M.~F., 
Cutri, R.~M., Stiening, R., et al.\ 2006, \aj, 131, 1163 

\bibitem[Soubiran et al.(2018)]{soubiran18} Soubiran, C., Cantat-Gaudin, T., Romero-G{\'o}mez, M., et al.\ 2018, \aap, 619, A155

\bibitem[Sung et al.(1999)]{sung} Sung, H., Bessell, M.~S., 
Lee, H.-W., Kang, Y.~H., \& Lee, S.-W.\ 1999, \mnras, 310, 982 

\bibitem[Tang et al.(2017)]{tang17} Tang, B., Geisler, D., Friel, E., et al.\ 2017, \aap, 601, A56. doi:10.1051/0004-6361/201629883

\bibitem[Taylor(2005)]{topcat} Taylor, M.~B.\ 2005, Astronomical Data Analysis Software and Systems XIV, 347, 29

\bibitem[Venuti et al.(2018)]{venuti18} Venuti, L., Prisinzano, L., Sacco, G.~G., et al.\ 2018, \aap, 609, A10. doi:10.1051/0004-6361/201731103

\bibitem[Wenger et al.(2000)]{wenger} Wenger, M., et al., 2000, \aaps, 143, 9

\bibitem[Wright et al.(2019)]{wright19} Wright, N.~J., Jeffries, R.~D., Jackson, R.~J., et al.\ 2019, \mnras, 486, 2477. doi:10.1093/mnras/stz870

\bibitem[Zacharias et al.(2013)]{zacharias} Zacharias, N., Finch, 
C.~T., Girard, T.~M., et al.\ 2013, \aj, 145, 44 


\end{thebibliography}
\end{document}